\documentclass[usenatbib]{mn2e}
\pdfoutput=1

\usepackage{newtxtext,newtxmath}
\usepackage[T1]{fontenc}
\usepackage{ae,aecompl}
\usepackage{docmute}
\usepackage{float}
\usepackage{comment}
\usepackage{color}
\usepackage{caption}
\usepackage{multirow}
\usepackage{docmute}
\usepackage{import}
\usepackage{graphicx}	
\usepackage{amsmath}	
\usepackage{amsfonts}
\usepackage{booktabs}
\usepackage{siunitx}
\usepackage[dvipsnames]{xcolor}
\usepackage{siunitx}
\usepackage{pifont}
\usepackage{tikz,hyperref}
\usepackage{aas_macros}
\usepackage{enumitem}

\newcommand{\tng}{TNG100}
\newcommand{\tngdm}{TNG100$_{\rm DM}$}
\newcommand{\fbxdm}{FIREbox$_{\rm DM}$}

\newcommand{\rockstar}{{\footnotesize ROCKSTAR}}
\newcommand{\subfind}{{\footnotesize SUBFIND}}
\newcommand{\mNine}{$10^{9}\ M_{\odot}$}
\newcommand{\mEight}{$10^{8}\ M_{\odot}$}
\newcommand{\mSeven}{$10^{7}\ M_{\odot}$}
\newcommand{\ltng}{$L_{\rm box}^{\rm TNG}$}
\newcommand{\lfbx}{$L_{\rm box}^{\rm FIRE}$}

\hypersetup{
  linkcolor  = blue,
  citecolor  = blue,
  urlcolor   = magenta,
  colorlinks = true,
}

\title[Out of sight, out of mind?]{Out of sight, out of mind? The impact of correlated clustering in substructure lensing}

\author[A. Lazar et al.]{Alexandres Lazar$^{1}$\thanks{\href{mailto:aalazar@uci.edu}{aalazar@uci.edu}},
James S. Bullock$^{1}$, 
Michael Boylan-Kolchin$^{2}$,
Robert Feldmann$^{3}$, \newauthor
Onur \c{C}atmabacak$^{3}$,
and Leonidas Moustakas$^{4}$ 
\\
$^{1}$Department of Physics and Astronomy, University of California, Irvine, CA 92697 USA\\
$^{2}$Department of Astronomy, The University of Texas at Austin, 2515 Speedway, Stop C1400, Austin, Texas 78712-1205, USA\\
$^{3}$Institute for Computational Science, University of Zurich, Zurich CH-8057, Switzerland\\
$^{4}$Jet Propulsion Laboratory, California Institute of Technology, 4800 Oak Grove Dr., Pasadena CA 91109, USA
\vspace{-0.4cm}
}

\date{Working Draft\vspace{-0.6cm}}
\pubyear{2020}

\begin{document}
\label{firstpage}
\pagerange{\pageref{firstpage}--\pageref{lastpage}}
\maketitle
\raggedbottom

\begin{abstract}
A promising route for revealing the existence of dark matter structures on mass scales smaller than the faintest galaxies is through their effect on strong gravitational lenses. We examine the role of local, lens-proximate clustering in boosting the lensing probability relative to contributions from substructure and unclustered line-of-sight (LOS) halos. Using two cosmological simulations that can resolve halo masses of $M_{\rm halo} \simeq 10^{9}\ M_{\odot}$ (in a simulation box of length $L_{\rm box}{\sim}100\,{\rm Mpc}$) and $10^{7}\ M_{\odot}$ ($L_{\rm box}\sim20\,{\rm Mpc}$), we demonstrate that clustering in the vicinity of the lens host produces a clear enhancement relative to an assumption of unclustered halos that persists to $> 20\,R_{\rm vir}$. This enhancement exceeds estimates that use a two-halo term to account for clustering, particularly within $2-5\,R_{\rm vir}$. We provide an analytic expression for this excess, clustered contribution.  We find that local clustering boosts the expected count of $10^9 \ M_\odot$ perturbing halos by ${\sim}35\%$ compared to substructure alone, a result that will significantly enhance expected signals for low-redshift ($z_l \simeq 0.2$) lenses, where substructure contributes substantially compared to LOS halos. We also find that
the orientation of the lens with respect to the line of sight (e.g., whether the line of sight passes through the major axis of the lens) can also have a significant effect on the lensing signal, boosting counts by an additional $\sim 50\%$ compared to a random orientations. This could be important if discovered lenses are biased to be oriented along their principal axis. 
\end{abstract}


\section{Introduction}
The large-scale clustering of galaxies provides important constraints on the makeup and evolution of the Universe \citep[e.g.][]{geller1989mapping,bond1996filaments,tegmark2004sdss,sanchez2006final}. The dark energy plus cold dark matter (CDM) paradigm, $\Lambda$CDM, is entrenched as the benchmark model for the theory of galaxy formation based largely on its success in matching observed large-scale structure.  For years, cosmological $N$-body simulations that incorporate only gravitational dynamics (dark matter only, DMO, simulations) have served as crucial tools for understanding the $\Lambda$CDM model, and have been used to understand the detailed  clustering of galaxies. When introducing full galaxy formation physics, cosmological simulations are able to match observed clustering statistics as a function of galaxy type as well \citep[e.g.][]{vogelsberger2014illustris,genel2014illustris,schaye2015eagle,khandai2015massiveblack,dubois2016horizon,dolag2016pathfinder,springel2018first,vogelsberger2020sims} 
however some discrepancies on smaller scales exist and motivate the exploration of alternative models \citep{bullock2017small,meneghetti2020excess}.

A feature of CDM that profoundly separates it from many other dark matter models is that CDM predicts a rich abundance of low-mass dark matter halos $M_{\rm halo} < 10^{6}\ M_{\odot}$ \citep{press1974formation,bullock2017small}. In cosmologies that include warm dark matter (WDM), for example, the  power spectrum is suppressed on scales smaller than a value set by the WDM particle mass \citep{bode2001wdm,bozek2016resonant}. For a thermal WDM particle of mass $m_{\rm thm} \lesssim 5\ \rm kev$, the formation of halos $< 10^{7}\ M_{\odot}$ \citep[e.g.][]{Schneider2013,horiuchi2016properties} is suppressed. Therefore, if halos below $\simeq 10^{7}\ M_{\odot}$ are detected, this would impose significant constraints on both the dark matter power spectrum and the particle nature of dark matter \citep[see][]{bertone2018era}.

Dark matter halos of sufficiently low mass are expected to be unable to form stars or retain baryons in the presence of a cosmological photoionizing background \citep[e.g.][]{efstathiou1992,bullock2000}. The detection of these starless halos, {\em with the abundance and density structure predicted by simulations}, would be triumphant for the CDM model. One way of inferring the presence of these low-mass objects is by their influence on cold, low-velocity stellar streams in the Milky Way \citep{ibata2002streams,calberg2009streams,yoon2011streams}. Recently \cite{banik2019novel} argued that the observed perturbations of the MW's stellar streams can only be explained by a population of subhalos in CDM. They set constraints to alternative dark matter models for halos down the mass function, notably setting a lower limit on the mass of warm dark matter thermal relics $m_{\rm thm} \gtrsim [4.6-6.3]\ \rm keV$. In order to provide tighter constraints for substructure down to $M_{\rm halo} \simeq 10^{5-6} M_{\odot}$ populating the MW, a larger sample cold streams would be needed. Another proposed approach to detecting these low-mass halos could be through the kinematics of stars in the Milky Way's disk \citep{feldmann2015gaia}.

Currently, the field of strong gravitational lensing offers to be another tool for the indirect detection of low-mass, starless halos of masses $\simeq 10^{6-8}\ M_{\odot}$  \citep{dalal2002direct,koopmans2005imaging,vegetii2010detection,vegetti2014inference,li2016constraints,nierenberg2017probing}. Lensing perturbations can arise from both subhalos within the lens host and from small halos found outside of the virial radius that perturb the light from source to the observer \citep[dubbed ``line-of-sight'' (LOS) halos;][]{li2017lens,despali2018modelling}. Notably, the field of substructure lensing offers tantalizing prospects, as in the near future, we expect both a gross increase in the number of lenses as well as a boost in resolution for instrument sensitivity. Forecasts suggest that the Dark Energy Survey (DES), LSST and EUCLID should discover hundreds of thousands of galaxy-galaxy lensing systems \citep{collett2015forthcoming}. The Nancy Grace Roman Space Telescope (RST) also has the potential of providing complementary catalogs of lens images \citep{weiner2020RST}. Additionally, the detection of halos with $M_{\rm halo} \sim 10^{7-8}\ M_{\odot}$ might be possible with JWST via quasar flux ratio anomalies \citep{macleod2013detection}. As of now, reported detections using ALMA reach the mass scale of classical Milky Way satellites ($M_{\rm halo} \sim 10^{10} M_{\odot}$), with constraints on subhalos an order of magnitude smaller. In the future, further observations and improved constraints may significantly improve these limits \citep{hezaveh2016detection}
and offer tighter constraints on the warm dark matter mass  \citep{he2020forward}, especially when combined with Milky Way satellite constraints \citep{nadler2021lensing}.

The expected count of subhalos that exist within the virial radius of the lens system has been studied rigorously. Studies of subhalo populations and their effect on lensing have previously relied on DMO simulations \citep[e.g.][]{bradac2002mass,xu2009aquarius,xu2012effects,metcalf2012small,vegetti2014inference}. More recently, however, simulations that implement full galaxy formation physics show that small subhalos are actually depleted with respect to DMO simulations, owing to interactions with the central galaxy   \citep{brooks2014baryons,wetzel2016reconciling,zhu2016baryonic,graus2018smooth,kelley2019elvis,richings2021lens}. Notably, \cite{sgk2017lumpy} showed that it is the central galaxy potential itself, not feedback, that drives most of the factor of $\sim 2$ difference in subhalo counts between DMO and full physics simulations for Milky Way-mass halos ($M_{\rm vir} \simeq 10^{12}\ M_{\odot}$). For lens-mass halos of interest ($M_{\rm vir} \simeq 10^{13}\ M_{\odot}$), \cite{despali2017impact} used both DMO and full physics from the EAGLE \citep{schaye2015eagle} and Illustris \citep{vogelsberger2014illustris} simulations to investigate predictions for subhalo lensing and found that simulations with full galaxy formation physics reduces the average expected substructure counts. The substructure analysis done in \cite{graus2018smooth} using Illustris found similar results.

Given the expected depletion in subhalo counts seen in full-physics simulations, the contribution of lensing signals from LOS halos has been recognized as ever more important. If LOS halos dominate the signal from a given lens, then uncertainties are reduced substantially because the contribution of the LOS halos can be accurately calculated, independent of the effect of baryonic physics, for a variety of cosmologies. Efforts have been made to understand the contribution of the LOS structure on the flux-ratio anomalies in lensed quasars \citep[e.g.][]{metcalf2005importance,metcalf2012small,xu2012effects,inoue2012weak,inoue2015constraints,xu2015flux,inoue2016origin,gilman2018forward,gilman2019flux}. It has become increasingly apparent that LOS halos should dominate the signal for more distant lenses ($z_l \sim 0.5$) while the contribution from subhalos should be non-negligible for more local lenses ($z_l \sim 0.2$; \citealt{despali2017impact}). 

Our objective in this article is to understand and quantify an effect not discussed in most previous work: how does local, correlated structure, in the vicinity of the lens host halo, impact the expected lensing signal?
While such an effect has been discussed before in the context of weak lensing \citep{daloisio2014effect}, its impact in strong lensing remains elusive.
This effect will be most important for low-redshift lenses, where subhalos are known to contribute non-trivially compared to the LOS count. 
We use a suite cosmological simulations, including those that include both DMO and and full galaxy formation physics, to explore this question. Specifically, we quantify correlated structure in lens-centered projections of targets lens halos of $M_{\rm vir}\simeq 10^{13}\,M_{\odot}$ at redshift $z=0.2$, corresponding with the benchmark sample discussed in \cite{vegetti2014inference}. This is done using two simulation projects: The first, from the public IllustrisTNG project \citep{nelson2017first}, includes both DMO and full-physics versions and resolves halos down to $M_{\rm halo} = 10^{9}\,M_{\odot}$ in a fairly large cosmological volume. The second is a DMO simulation evolved in much smaller cosmological volume that is able resolve dark matter halos down to $ M_{\rm halo} = 10^{7}\, M_{\odot}$. 

This paper is structured as follows. Section~\ref{sec:numerical.methods} introduces our set of simulations, provides a description of the selected sample of halos, and outlines our methodology for counting structures along projected line-of-sights in the simulations. Section~\ref{sec:results} provides results on structure and explores how the lens-host orientation can boost the amount of structure expected along lens-centered projections. Finally, we summarize our results in Section~\ref{sec:conclusion}. 
Note that in strong lensing literature, terms such as ``field halo'' and ``line-of-sight (LOS) halo'' typically refer to the same things. We will use these terms interchangeably throughout this work to refer to halos having a volume density equal to the cosmological mean density of halos at that mass.

\section{Numerical Methodology}
\label{sec:numerical.methods}
Our $\Lambda$CDM predictions rely on two sets of simulations. The primary set comes from the public catalogs of the IllustrisTNG project\footnote{The Illustris data is publicly available at \url{https://www.tng-project.org/}} \citep{nelson2017first,marinacci2018first, springel2018first, naiman2018first,pillepich2017first}. As described in Section~\ref{sec:TNG}, these simulations allow us to identify halo populations robustly to masses greater than $10^{9}\ M_{\odot}$ for a {\em large-scale} environment both with gravitational physics alone and with full galaxy formation physics. The second simulation, introduced in Section~\ref{sec:FIRE}, is a DMO version of a new simulation suite called FIREbox, which is part of the Feedback in Realistic Environments (FIRE) project.\footnote{The {\small FIRE} project website: \url{ http://fire.northwestern.edu}} 
The mass functions of these simulations are presented in Fig.~\ref{fig:1}. Both simulations assume a Planck 2015 cosmology \citep{ade2016planck}: $\Omega_{\rm m} = 0.3089$, $\Omega_{\Lambda} = 0.6911$, $\Omega_{\rm b} = 0.0486$, $\sigma_{8} = 0.8159$, and $h = 0.6774$.

Our analysis relies on halo catalogs at redshift $z=0.2$, which coincides with a typical redshift of the lens-host sample explored in \cite{vegetti2014inference}. Dark matter halos are defined to be spherical systems with a virial radius, $R_{\rm vir}$, inside which the density is equal to the average density of $\Delta_{\rm vir}(z)\rho_{\rm crit}(z)$, where $\Delta_{\rm vir}(z)$ is the virial overdensity defined by \cite{bryan1998statistical} and $\rho_{\rm crit}(z)$ is the
critical density of the Universe. The virial mass, $M_{\rm vir}$, is then the total mass enclosed in a sphere of radius $R_{\rm vir}$. 

In what follows, we discuss two types of halos: First, we have massive {\em target} halos, chosen to mimic lens-galaxy hosts: $M_{\rm vir} \in [0.8 - 2] \times 10^{13}\ M_{\odot}$. Second, we have low-mass perturber halos, which can either be subhalos or small halos that exist somewhere outside of the host's virial radius and along the LOS from the observer. We quote their masses (taken from the halo catalogs) using the symbol $M_{\rm halo}$, since for subhalos $M_{\rm vir}$ is not physically relevant.

\subsection{The IllustrisTNG simulations}
\label{sec:TNG}
The IllustrisTNG (TNG) suite of cosmological simulations was run using the moving-mesh code {\footnotesize Arepo} \citep{springel2010arepo}. The runs with full galaxy formation physics use an updated version of the Illustris model \citep{weinberger2017tng,pillepich2017first}. We use the high resolution set of publicly-available simulations,\footnote{The highest resolution box is actually TNG50-1, but was not publicly available by the time this article was submitted.} TNG100-1, which has a comoving box of side length $L_{\rm box}^{\rm TNG}=75\, h^{-1} \mathrm{cMpc}=110.7\,{\rm cMpc}$. The full physics run, \tng{}, has a dark matter particle mass of $m_{\rm dm} = 7.5 \times 10^{6}\ M_{\odot}$ and a gas particle mass of $m_{\rm gas}= 1.4 \times 10^{6}\ M_{\odot}$. The high-resolution simulation that uses dark matter only (DMO) physics, \tngdm, has the same box size, but treats baryonic matter as collisionless particles, in which gives the simulation a particles mass is $m_{\rm dm} = 8.9 \times 10^{6}\ M_{\odot}$. When comparing between \tng{} and \tngdm, we account for the excess baryonic mass in the DMO simulation by introducing a factor $m_{\rm dm} \rightarrow (1-f_{b})m_{\rm dm}$, where $f_{b} := \Omega_{\rm b}/\Omega_{\rm m}$ is the cosmic baryon fraction.

\begin{figure}
    \centering
    \includegraphics[width=\columnwidth]{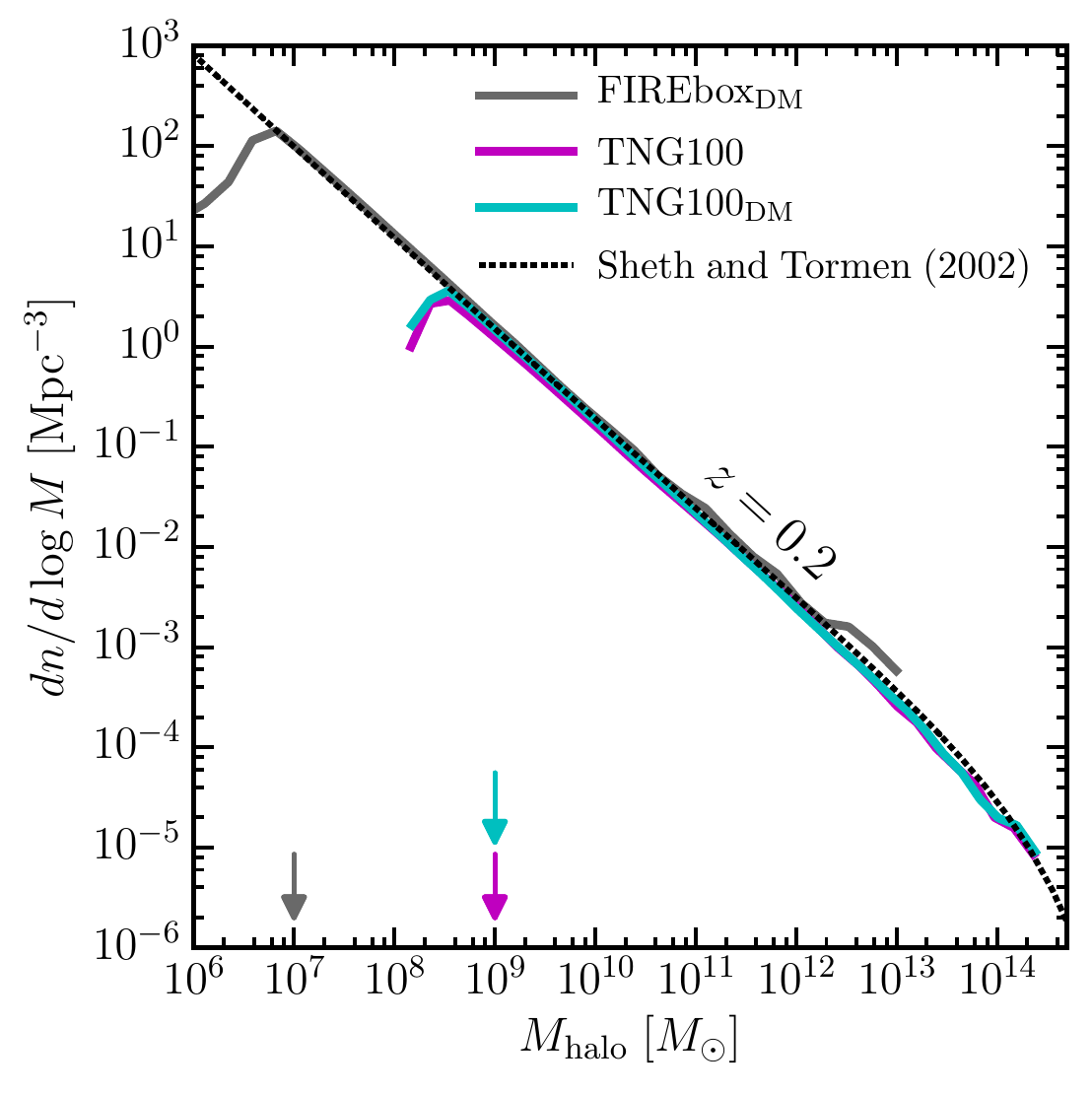}
    \caption{---
        {\bf \em The abundance of resolved dark matter halos}. Shown are the halo mass functions of the simulations at $z=0.2$.
        Results from all three simulation are in agreement with the analytical prediction of \protect\cite{sheth2002excursion}. The arrows pointing to the $x$-axis designate the minimally resolved mass considered in each simulation, indicated by the color coding.
    }
    \label{fig:1}
\end{figure}

For both \tng{} and \tngdm{}, the dark matter halo catalogs were constructed using a friends-of-friends ({\footnotesize FOF}) algorithm \citep{davis1985cdm} with a linking length $0.2$ times the mean inter-particle spacing. Gravitationally bound substructures are identified through the \subfind{} halo finder \protect\citep{springel2001populating}. These subhalos have a dark matter mass that is gravitationally bound to itself but not to any other subhalos found in the same FOF host or the host itself. 
As shown in Fig.~\ref{fig:1}, the $z=0.2$ mass functions of the \tng{} (magenta) and \tngdm{} (cyan) \subfind{} subhalos match well with the analytical \cite{sheth2002excursion} mass function (dotted dashed) at that redshift. Both simulations become incomplete just below $M_{\rm halo} \simeq 10^{8.6}\ M_{\odot}$.  In order to be conservative, we restrict our analysis to subhalos and field halos with resolved masses  $M_{\rm halo} > 10^{9}\ M_{\odot}$.  These halos contain at least 134 dark matter particles.   
\tng{} and \tngdm{} have 166 and 168 lens-target halos, respectively, where in \tng{}, these halos typically host galaxies with $M_{\star} \simeq 10^{11}\ M_{\odot}$, which is consistent with the benchmark sample in \cite{vegetti2014inference}.

The public \subfind{} catalogs in \tng{} include several baryon-dominated ``subhalos," many of which contain no dark matter. Most of these baryon-dominated objects exist within ${\sim} 20$ kpc of host galaxies and appear to be baryonic fragments numerically identified by \subfind{} rather than galaxies associated with dark matter subhalos. While baryonic clumps could induce perturbations detectable in lens images \citep{gilman2017strong,hsueh2017sharpiv,hsueh2018flux,he2018globular}, this type of object is not the focus of our analysis; we are interested in the search for low-mass {\em dark matter} structures.  For this reason, we exclude systems with a ratio of total baryonic mass to dark matter mass that is more that twice the cosmic baryon fraction, ($M_{\rm bar}/M_{\rm halo} > 2\times f_{\rm b}$) in the substructure analysis that follows. 

\subsection{The FIREbox DMO simulation}
\label{sec:FIRE}
FIREBox is a new effort within the FIRE collaboration to simulate cosmological volumes of $L_{\rm box}^{\rm FIRE} = 15\ h^{-1}\rm cMpc = 22.14\ cMpc$ at the resolution of FIRE zoom-in simulations (Feldmann et al., in prep). 
Our analysis uses the results from a DMO version of the FIREbox initial conditions, ran with $2048^{3}$ particles, which we dub \fbxdm{} in this article. \fbxdm{} has a dark matter particle mass of $m_{\rm DM}= 5.0 \times 10^{4}\ M_{\odot}$ and was run with a Plummer softening length of $\epsilon_{\rm DM} = 40\ \rm pc$.
Since we will work with a DMO simulation to complement the analysis done in the two TNG simulations, we again account for baryonic mass by multiplying a by conversion factor of $(1-f_{\rm b})$, as we do for \tngdm{}.These simulations are complete, conservatively, to halo masses down to $10^{7}\ M_{\odot}$ (see Fig.~\ref{fig:1}).

The halo catalog for \fbxdm{} was generated using \rockstar{} \citep{behroozi2012rockstar}. This method finds dark matter halos using a hierarchically adaptive refinement of {\footnotesize FOF} in a 6-dimensional phase-space and one time dimension. We set our \rockstar{} halo finding parameters to be comparable to those use in the TNG catalogs generated by \subfind. Specifically, we use a {\footnotesize FOF} linking length of $b=0.20$ and include only halos that have at least $100$ dark matter particles. We also set the criteria for unbound particle fraction rejection to $70\%$ instead of the default $50\%$, as explored in the Appendix of \cite{graus2018smooth}. Doing so minimizes ambiguities associated with using different halo finders for computing halo masses.\footnote{We have also experimented with higher unbound fractions of $90\%$ and $95\%$ with a fixed value of $b=0.20$. We chose the unbound fraction of $70\%$ and a {\footnotesize FOF} linking of $b=0.20$ because they provide the best match with \tngdm{} catalogs for $M_{\rm halo} > 10^{9}\ M_{\odot}$.}
With these choices, FIREbox contains three target halos with mass $M_{\rm vir} \in [0.8 - 2] \times 10^{13}\ M_{\odot}$. 

Returning to Fig.~\ref{fig:1}, the mass function of \fbxdm{} at $z=0.2$ is shown as the solid gray curve and  agrees well with the analytical mass function down to halo masses with $10^{7-8}\ M_{\odot}$. With this in mind, the use of \fbxdm{} in our analysis will be restricted to two sets of subhalo masses: a sample of halo masses down to \mSeven{} (${\sim} 10^{2}$ particles) and a sample of down to \mEight{} (${\sim} 10^{3}$ particles). For a more stringent check, Appendix~\ref{sec:compare} compares the subhalo mass function and the subhalo $V_{\rm max}$ function of the three target-lens halos of \fbxdm{} to the same mass target-lens halos in \tngdm{}. We find excellent agreement between substructure statistics between our halo samples. 

\begin{figure}
    \centering
    \includegraphics[width=0.925\columnwidth]{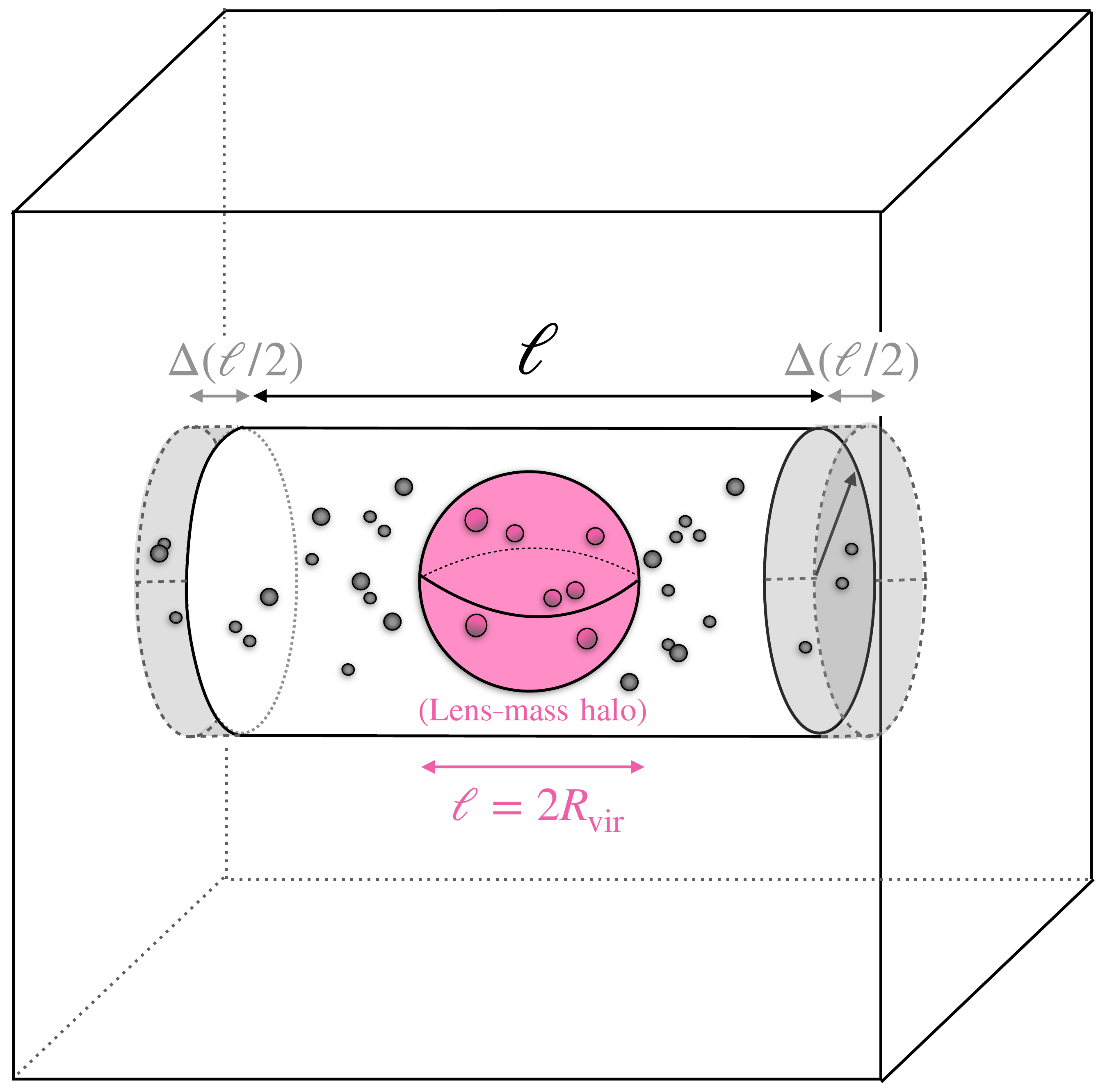}
    \caption{
    A cartoon depiction of the analysis we perform for each lens-centered cylinder in a simulation box. The lens-mass halo, centered at the origin of the box, is illustrated as the large pink sphere, while subhalos are illustrated as smaller halos. With the cylinder held at a fixed $\mathcal{R}$, the cylinder is varied by increments of $\Delta{\ell}$ ($\Delta\ell/2$ at each end of the cylinder) until reaching the edge of the box. Note that the radius of the cylinder has been exaggerated for clarity.  In practice, $\mathcal{R} \ll R_{\rm vir}$.
    }
    \label{fig:2}
\end{figure}

\begin{figure*}
    \centering
    \includegraphics[width=0.925\textwidth]{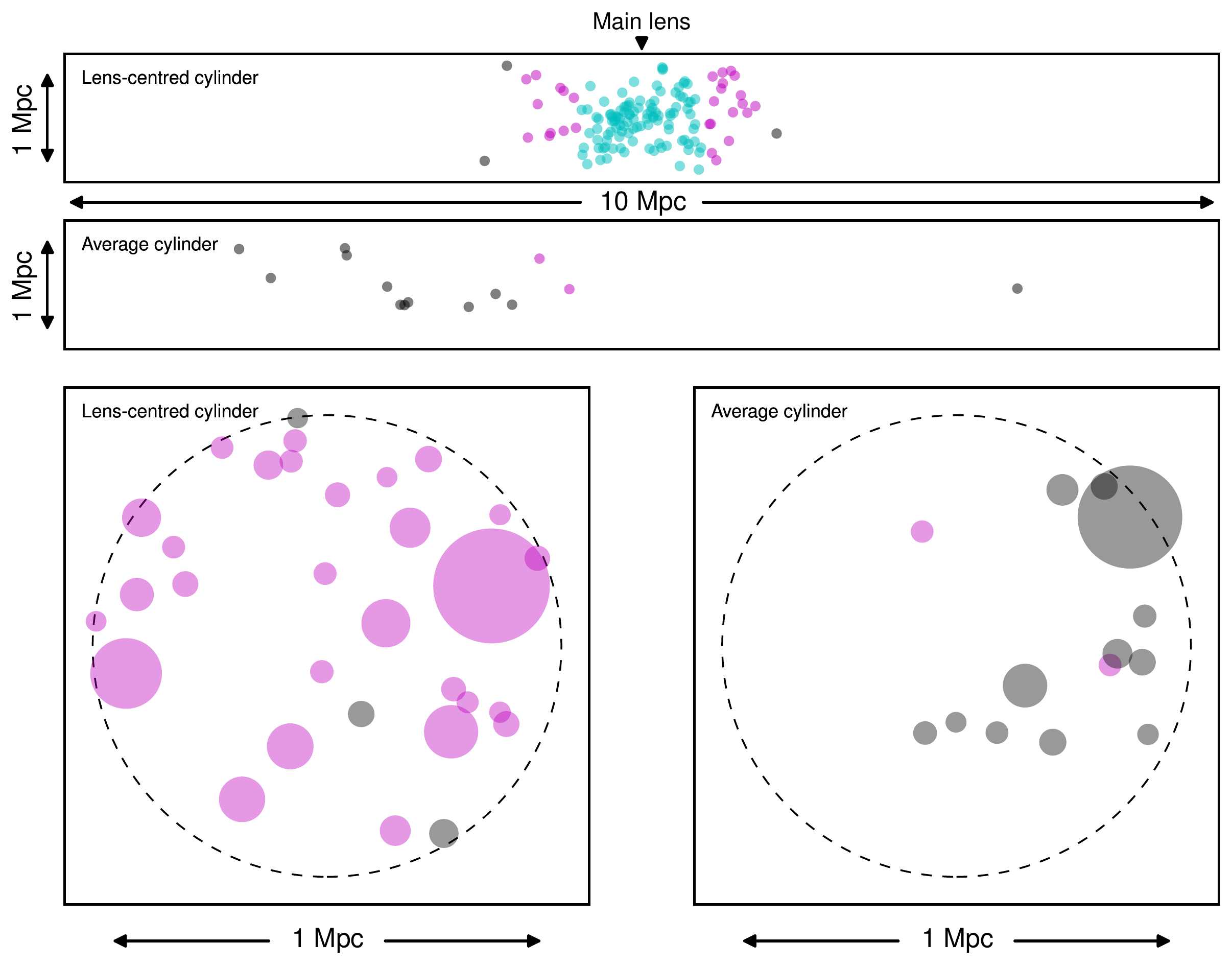}
    \caption{ ---
        {\bf \em The importance of correlated structure in sub-galactic lensing}. 
        The upper and middle panels depict the side-view of a cylinder length of $\ell = 10\ \rm Mpc$ and radius $\mathcal{R} = 500\ \rm kpc$ centered on a lens-mass host halo with radius $R_{\rm vir} \simeq 500\ \rm kpc$ at $z=0.2$ in the \tng{} simulation. The points show locations of $M_{\rm halo} > 10^{9}\ M_{\odot}$ halos within the cylinder, color coded by relative distance from the host out to the size of the halo $\ell=2R_{\rm vir}$ (or equivalently to $r=R_{\rm vir}$; cyan points), within $\ell=[2-4]R_{\rm vir}$ ($r=[1-2]R_{\rm vir}$; magenta points), and everything else outside $\ell=4R_{\rm vir}$ out to $\ell=10$ Mpc (black points). The figure beneath it shows an identical cylinder that samples a representative region of the simulation box using the same color scheme; the area of each point is proportional to that halo's mass. The square plots are the same cylinders shown face on, centered on the lens (left) and centered randomly (right). The cylinder radius $\mathcal{R}=500\,{\rm kpc}$ used for this figure is, for illustrative purposes, much larger than the typical Einstein radius of the host ($\mathcal{R} \approx 10\,{\rm kpc}$; we use the latter value for our actual analysis. {\it Local, correlated perturbers around the lens are highly significant} (compare the magenta points in the left verses right panels on the bottom). 
    }
    \label{fig:3}
\end{figure*}

\begin{figure*}
    \centering
    \includegraphics[width=0.925\textwidth]{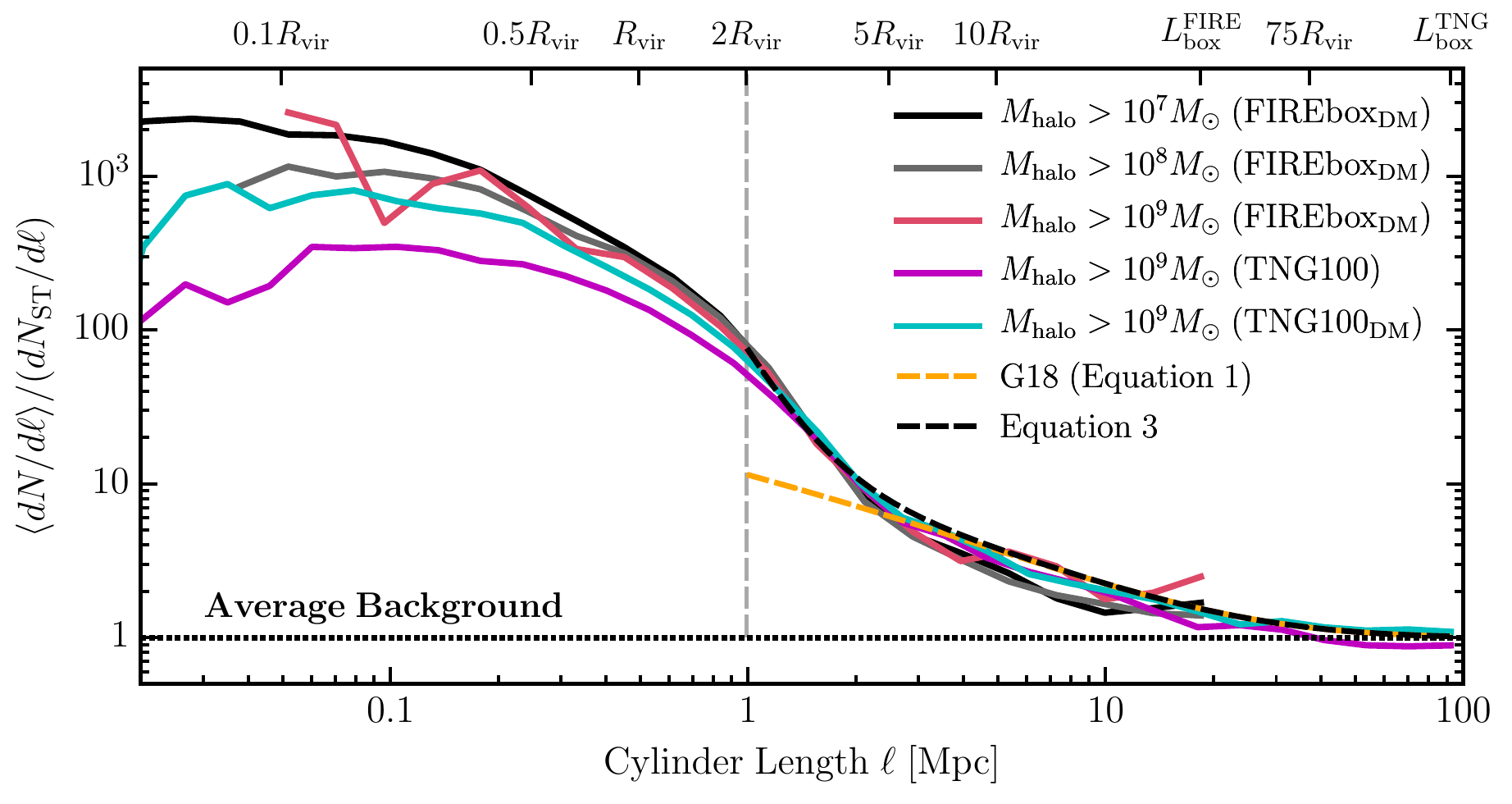}
    \caption{---
        {\bf \em Structure excess along lens-centered projections}:
        The mean differential count $\langle dN/d\ell \rangle$ of small halos within cylinders of increasing length $\ell + \Delta{\ell}$ centered on the lens-mass host halos ($M_{\rm vir} \simeq 10^{13} M_{\odot}$, solid colored), normalized by the expected average background counts via \protect\cite{sheth2002excursion} halo mass function (dotted black). The cylinders mirror those shown in Fig~\ref{fig:3}, but now with a radius comparable to the Einstein radius of the lens-mass halos, $\mathcal{R} = 10\ \rm kpc$ (and varying length $\ell$).
        The gray vertical dashed line marks the typical outer region of the halo $(\ell = 2R_{\rm vir} \leftrightarrow r=R_{\rm vir})$.
        Inside the lens-mass host halos ($\ell < 2 R_{\rm vir}$), the differential counts are self-similar for the DMO simulations. For \tng{} (solid magenta), the counts decrease by almost a factor of 10 because of the destructive effect of the central galaxy. Differences between the curves become apparent outside the halo due to local halo clustering. This effect originates predominantly from ``backsplash" halos that have pericenters with $r<R_{\rm vir}$ but apocenters of $r > R_{\rm vir}$, meaning they lie within the splashback radius of the lens-mass host but spend most of their time at $r > R_{\rm vir}$. 
        For comparison, we plot the analytical contribution of the two-halo term given by G18 (Eq.~\ref{eq:analytical.los}; dashed orange), where we set $\delta_{\rm los} = 1$. While G18 accurately reproduces the contribution at $\ell > 5\times R_{\rm vir}$ for most of the curves, it significantly under-predicts the differential halo counts within $\ell \in [2-5] R_{\rm vir}$. Our proposed modification, Eq.~\eqref{eq:analytical.los.corrected} (dashed black), captures the excess for $\ell > 2\,R_{\rm vir}$ to within $10\%$.
    }
    \label{fig:4}
\end{figure*}

\subsection{Counting within lens-centered cylinders}
\label{sec:orienation}
We quantify perturber counts by enumerating small halos that sit within randomly-oriented {\em lens-centered cylinders} of length $\ell$ and radius $\mathcal{R}$, where $\ell$ can extend over the length of the simulation box and $\mathcal{R}$ will be set to a value close to an expected Einstein radius ($\sim 10$ kpc). As illustrated in 
Fig.~\ref{fig:2}, each cylinder is centered on a target halo in the simulation box. The volume of the cylinder is increased by lengthening $\ell \rightarrow \ell+\Delta\ell$ with the radius $\mathcal{R}$ fixed. For a discrete increment variation of $\Delta{\ell}$, both ends of the cylinder are increased by $(\Delta\ell)/2$, which captures both structures whose positions are found in the foreground and background of the target halo. Doing so allows us to qualitatively compare counts as function of radius $r$ from the halo center, i.e., $r\simeq\ell/2$. For example, a lens-centered cylinder length of $\ell = 2 R_{\rm vir}$ spans the full diameter of the dark matter halo. 

In what follows we compare counts from lens-centered cylinders to those of {\em average} cylinders, which are configured like a lens-centered cylinder, but now with their centers randomly placed within the simulation box. Note for a large sample of average cylinders, the mean count per unit volume at any $\ell$ will be equal to the average halo density \citep[e.g.][]{sheth2002excursion}. For both cylinder types, we take into account the periodicity of the cosmological box but never allow the full cylinder length to exceed the box length.

Fig.~\ref{fig:3} provides a simple illustration of our counting prescription applied to \tng{} for a single lens-mass halo of $M_{\rm vir} \simeq 10^{13} M_{\odot}$ (labeled as ``lens-centered cylinder'') and for a randomly placed cylinder of the same size inside the box (labeled as ``average cylinder'').For illustrative purposes, we have set the radius of the cylinder to a very large value $\mathcal{R} = R_{\rm vir} \simeq 500\ \rm kpc$, much larger than what we will use in our main analysis. The cylinder length is set to $\mathcal{\ell}=10$ Mpc. The filled circles show all subhalos with $M_{\rm halo} > 10^{9}\ M_{\odot}$. Points and circles are colored based on their relative distance from the host: within $\ell=2R_{\rm vir}$ ($r=R_{\rm vir}$) as cyan, $\ell=[2-4]R_{\rm vir}$ ($r=[1-2]R_{\rm vir}$) as magenta, and black for everything else out to a length $\ell=10$ Mpc. The two top plots show the edge-on projections while the bottom plots are the face-on projections. For the face-on projections, the sizes of the circles are scaled proportionally to the mass of the halos in the cylinder (with the host halo removed for clarity). In the bottom-left panel, we show only halos that are {\em outside} of $R_{\rm vir}$ of the lens-host. Even excluding substructure, the overall count is much higher than the random cylinder.

For this particular randomly-chosen "average" cylinder,  we find 14 halos with $M_{\rm halo} > 10^{9}\ M_{\odot}$ within the 10 Mpc projection visualized. Note that the cosmological average expected for this volume is $\approx 14.3$ when using the \cite{sheth2002excursion} mass function. 
Counts are significantly higher for the lens-centered cylinder. For comparison, the lens-centered cylinder contains 108 halos of the same mass. Interestingly, correlated structure counts around the lens-host that exist outside of $r=R_{\rm vir}$ but within $r=2 R_{\rm vir}$ is 28, which exceeds all of the counts within the 10 Mpc long average cylinder. This shows that clustering in the vicinity of the lens will boost signals non-trivially compared to what we would have estimated by ignoring local clustering outside of the halo virial radius. 

While Fig.~\ref{fig:3} is useful to elucidate the point of  a projected cylinder, the radius shown, $\mathcal{R} = R_{\rm vir}$, is not relevant for lensing studies. The remainder of the analysis hereafter imposes a fixed projected cylinder radius of $\mathcal{R} = 10\ \rm kpc$, which is a value comparable to size of the lens-mass' Einstein radius (typically $\lesssim$ 10 kpc). 
Note that in the SLACS sample used by \cite{vegetti2014inference} at a redshift $\langle z_{\rm lens} \rangle\sim 0.2$, the median Einstein radius is $\sim 4.2$ kpc. We adopt a slightly larger 10 kpc radius in what follows in order to improve counting statistics (see Appendix~\ref{sec:dmo.sampling} for a discussion of counting variance).  Our primary results below are framed as relative counts per unit volume, such that the precise radius of the cylinder factors out. We show in Appendix~\ref{sec:cylinder.choice} that our results are insensitive (to within counting noise) to choices of cylinder radii of $5$ kpc and even $2$ kpc.

\section{Results}
\label{sec:results}
\subsection{Average halo counts in projection}
\label{sec:rate.of.counts}
Our main results are presented in Fig.~\ref{fig:4}, where we plot the {\em mean} differential count of halos per cylinder length, $\langle dN/d\ell\rangle$ using 100 cylinders randomly oriented around each lens-mass halo (solid lines). Integrating $dN/d\ell$ over $\ell$ gives the cumulative count $N(<\ell)$ within a cylinder of total length $\ell$. We plot the differential count as a function of cylinder length, $\ell$.  Shown is the {\it mean} differential count rather than the {\em median} because this allows us to compare directly to analytic expectations for the average halo abundance.
Each solid curve shows the rate of counts for halo masses greater than a given value (indicated in the legend), normalized by the rate of counts expected from the average background from \citep{sheth2002excursion}. Counts equal to the rate of counts from the average background are shown as the horizontal dotted line with an amplitude of $1$.\footnote{We have also tested with a large number average cylinders in the simulations (as the example shown in Fig.~\protect\ref{fig:3}) and confirmed that the average background counts are consistent with analytical expectations of \protect\cite{sheth2002excursion}.} The vertical grey-dashed line separates between the two regimes of interest: the substructure contribution $(\ell < 2 R_{\rm vir}$ and the local structure $(\ell > 2 R_{\rm vir})$.

We see that for all simulations and mass cuts, the differential counts are above the average counts out to $r \sim 20$ Mpc ($\ell\sim 40\,R_{\rm vir}$). This is attributed to excess clustering in the vicinity of the massive host, an effect often ignored in lensing studies \citep[see e.g.][who assume average counts outside the virial radius]{despali2018modelling}, though some groups \cite[e.g.][]{gilman2018forward} have attempted to account for the effect (see below). For perturbers more massive than $10^{9}\ M_{\odot}$, the rate of counts do not reach the average background until $\ell \sim 75 R_{\rm vir} \approx 40\ \rm Mpc$ for both \tng{} (magenta), \tngdm{} (cyan), and \fbxdm{} (red, which is limited by only having three hosts).  It is interesting also to compare \tng{} (magenta) to \tngdm{} (cyan).  We see that at small $\ell$ (corresponding to the center of the host halo), the overall count is higher in the DMO run.  This comes about because of enhanced subhalo destruction from the central galaxy potential \citep[e.g.][]{graus2018smooth}. 
Differential counts in \fbxdm{} are consistent with those in \tngdm{} for a lower mass threshold of $M_{\rm halo}= 10^{9}\, M_{\odot}$, though the \fbxdm{} result is noisier owing to the fact that there are only three target lens-mass halos in the volume. In Appendix~\ref{sec:dmo.sampling}, we further discuss the effect of sample variance in our analysis. 

Comparing the \mNine{} (magenta/cyan), \mEight{} (gray) and \mSeven{} (black) lines, there is an indication that lower-mass halos contribute more near the center of the lens-host ($\ell \sim 0.1 R_{\rm vir}$).  This would be expected if subhalo radial distributions within the host halo are more centrally concentrated at lower subhalo masses.  
Beyond the virial radius $(\ell > 2\,R_{\rm vir})$ the lower-mass halos found in \fbxdm{}, \mEight{} (gray) and \mSeven{} (black), track the enhanced counts seen at  $10^{9}\ M_{\odot}$. For $\ell > 5R_{\rm vir}$, the counts for lower mass halos in \fbxdm{} fall slightly below those seen in TNG. This difference could be physical.  For example,  lower mass halos may be less clustered around the lens host.  However,  the offset we see from \mNine{} to \mEight{} is much larger than would be expected naively from the clustering bias change over this mass range: $b(10^9 M_\odot) \simeq 0.64$ vs. $b(10^8 M_\odot) \simeq 0.63$ at $z=0.2$ \citep{sheth1999bias}. The difference could also arise from the lack of large-scale power in the small volume of \fbxdm{} or from simple sample variance from having only three host halos. 
In Appendix~\ref{sec:dmo.sampling}, we demonstrate that these differences are consistent with sample variance; in order to definitively confirm this, we would need a larger cosmological box with a particle mass resolution comparable to \fbxdm.

\subsection{Analytic Model Comparison}
\label{sec:analytic}
Correlated structure outside of the virial radius of a massive target halo is related to the "two-halo term" of the halo-matter correlation function \citep[e.g.][]{ma2000stabilize,seljak2000analytical,smith2003clustering}.  \cite{gilman2018forward} (G18 hereafter) estimated this effect as
\begin{align}
    \frac{d^{2}N}{dm\ dV}
    &=
    \delta_{\rm los}
    \Big[ 1 + \xi_{\rm 2halo}(r,M_{\rm vir},z) \Big]\frac{d^{2}N_{\rm ST}}{dm\ dV},
    \, ,
    \label{eq:analytical.los}
\end{align}
where $M_{\rm vir}$ is the mass of the host halo and $\delta_{\rm los}$ is an overall scaling term that accounts for a systematic shift of the mean number of halos predicted by the \cite{sheth2002excursion} mass function ($N_{\rm ST}$), and
\begin{align}
    \xi_{\rm 2halo}(r, M_{\rm vir}, z)
    &=
    b_{\rm ST}(M_{\rm vir}, z)
    \xi_{\rm lin}(r,z)
\end{align}
is the two-halo term that depends on the bias, $b_{\rm ST}$, around the lens halo computed as in \cite{sheth1999bias} and $\xi_{\rm lin}$ is the linear matter-matter correlation function at a three-dimensional distance, $r$, computed from the linear power spectrum at redshift $z$. 

\begin{figure*}
    \centering
    \includegraphics[width=0.925\textwidth]{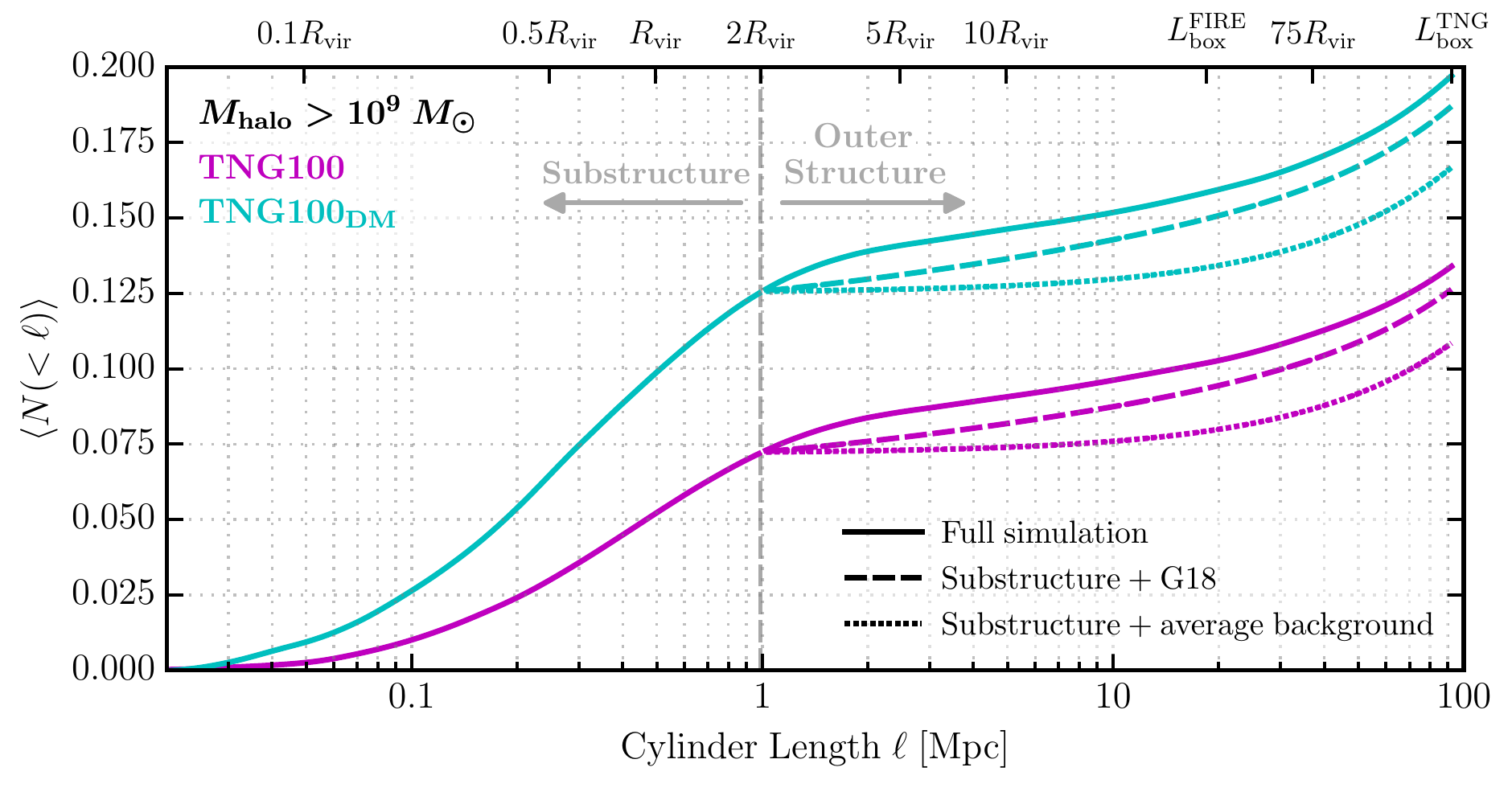}
    \caption{
        Solid lines show the integrated mean count of halos more massive than \mNine{} within cylinders of radius $\mathcal{R} = 10$~kpc and of length $\ell$ centered on lens hosts from \tng{} (magenta) and \tngdm{} (cyan). The gray vertical line marks the cylinder length at the halo boundary ($\ell = 2 R_{\rm vir}$). The cumulative count of halos that would result without any adjustment for correlated structure beyond the virial radius is shown by the dotted lines. The predicted contribution of correlated structure from G18  (Eq.~\ref{eq:analytical.los.corrected}) is shown by the dashed lines. 
         In implementing G18, we set $\delta_{\rm los} =1$ for \tngdm, as the convergence to average is about one-to-one. However, for \tng, we set $\delta_{\rm los} =0.879$ to remain consistent with the average-count differences in \tngdm. The G18 model captures much, but not all, of the correlated structure; the extra component beyond the G18 prediction comes from virialized halos beyond $R_{\rm vir}$ but within the splashback radius.
    }
    \label{fig:5}
\end{figure*}

Eq.~\eqref{eq:analytical.los} (labeled ``G18'') is plotted as the orange dashed curve in Fig.~\ref{fig:4}. There, we designate $\delta_{\rm los} = 1$ since the structure found in the volumes of all three simulations tend to be well represented by the halo mass function (refer to Fig.~\ref{fig:1}). Notice that in the region $\ell = [2-5]R_{\rm vir}$, counts in the simulation are in excess of the G18 estimate. This excess likely originates from subhalos with orbits that have apocenters beyond $R_{\rm vir}$. This ``backsplash" population can be substantial just outside of $R_{\rm vir}$, as 50-80\% of halos at $r \in [1-1.5]\,R_{\rm vir}$ were once subhalos \citep[e.g.][]{gill2005outskirts,sgk2014elvis} and therefore represent a natural continuation of the subhalo population within $R_{\rm vir}$\footnote{A backsplash excess is also hinted at pictorially in Fig.~\ref{fig:3} (top panel) from the magenta dots in the (edge-on) lens-centered  projection}. A physical boundary for this virialized population of halos is the so-called ``splashback" radius of the host \citep{more2015splash}, where recently-accreted material reaches its second apocenter (or the first apocenter after turn-around, where the turn-around -- or infall -- radius is $R_{\rm infall} \approx 1.4\,R_{\rm sp}$). 
Our sample of lens-mass halos at  $z=0.2$ should have a median splashback radius of $R_{\rm sp} \approx 1.5 R_{\rm vir}$ \citep{more2015splash}. Subhalos outside of $R_{\rm vir}$ but within $R_{\rm sp}$ ($\ell \approx 3 R_{\rm vir}$), and accompanying halos on first infall ($r< R_{\rm infall} \approx 2.1 \, R_{\rm vir} \leftrightarrow \ell \la 4.2 \,R_{\rm vir}$) therefore provide a natural explanation of the excess relative to G18 at $\ell = [2-5]R_{\rm vir}$.  Notably, the G18 model matches our simulation results for $\ell > 5R_{\rm vir}$, similar to the radius at which \cite{sgk2014elvis} find the backsplash fraction is essentially zero.

To accommodate the excess clustering seen within $5R_{\rm vir}$, we modify Eq.~\eqref{eq:analytical.los} as
\begin{align}
    \frac{d^{2}N_{\rm CDM}}{dm\ dV}
    &=
    \delta_{\rm los}
    \left[ 1 + \xi_{\rm 2halo}(r, M_{\rm vir}, z) + b_{\rm sp}(r) \right]\, \frac{d^{2}N_{\rm ST}}{dm\ dV}
    \,,
    \label{eq:analytical.los.corrected}
\end{align}
where 
\begin{align}
   b_{\rm sp}(r)
    &:=
    b_{e}\left( \frac{r}{5R_{\rm vir}}\right)^{-s_{e}}
    \,.
\end{align}
Here $b_{e}$ and $s_{e}$ are free parameters, and the term in parentheses accounts for the excess clustering (compared to Eq.~\ref{eq:analytical.los}). Eq.~\eqref{eq:analytical.los.corrected} explicitly separates the contribution of the cosmologically average LOS halos, an enhancement from halo-halo clustering, and a further enhancement from backsplash halos. The implied LOS count with this parametrization is shown as the black-dashed curve in Fig.~\ref{fig:4} with $b_{e}=0.1$ and $s_{e}=4$. This choice of parameters captures our results to an accuracy of  $10 \%$ all the way out to the edge of the simulation box. In particular, our parametrization is what we would expect for a population of virialized (sub)halos that populate between distances of $R_{\rm vir}$ and $5 R_{\rm vir}$ 

\subsection{Cumulative Counts}
\label{sec:counts}
In order to estimate sample size of suitable lenses needed for testing predictions, it is useful to know how many low-mass halos, on average, we expect to see along the LOS to a host.  The lower the expected count per lens, the more lenses we will need to place meaningful constraints on the halo mass function at low masses. The average count will depend, of course, on the redshift of the target galaxy and lens relative to the observer \citep[e.g.][]{despali2017impact} but our results allow us to determine the average count within the $\sim 100$ Mpc vicinity of the lens. Broadly speaking, the closer the lens to the observer, the more important correlated structure will be. {\em The results that follow will be important for any lens where the substructure contribution is significant compared to the total expected count of perturbers.} This is the case for roughly half of the lenses in the \citet{despali2017impact} sample, for example.

Fig.~\ref{fig:5} shows the mean count of halos along the projected lens-centered cylinder as a function of cylinder length $\ell$, where the mean cumulative counts, $\langle N(<\ell) \rangle$, is related to the average rate of counts by
\begin{align}
    \left\langle N(<\ell) \right\rangle
    =
    \int_{0}^{\ell}
    d\ell'\  \left\langle \frac{dN}{d\ell'} \right\rangle
    \, .
    \label{eq:avg.los}
\end{align}
Following the presentation of Fig.~\ref{fig:4}, for increasing $\ell$, the simulation results fully realize the expected clustering contribution to the LOS halo count once integrating the solid lines out to the cosmological boxes. 

Solid lines in Fig.~\ref{fig:5} show the integrated mean count of halos more massive than \mNine{} within cylinders of length $\ell$ centered on lens hosts from \tng{} (magenta) and \tngdm{} (cyan). The gray vertical line marks the cylinder length at the halo boundary ($\ell = 2 R_{\rm vir}$). As we can see by comparing \tng{} and \tngdm\ at $\ell < 2R_{\rm vir}$, subhalo counts are reduced by $\approx 40$ in the full-physics run compared to the DMO run. This result is consistent with the findings of \cite{despali2017impact} and \cite{graus2018smooth}.
A second takeaway from Fig.~\ref{fig:5} is that the average count of halos is much less than unity out to the edge of the box. This means that most of the lens-centered LOS cylinder {\it do not} contain halos larger than \mNine within within 100 Mpc in projections of radius $\mathcal{R} = 10$kpc. We find that for \tngdm, $87.5\%$ of projections contain no halos at this mass limit.  For \tng, the fraction of empty projections rises to $\approx 92.5\%$. 

\begin{figure*}
    \centering
    \includegraphics[width=0.925\textwidth]{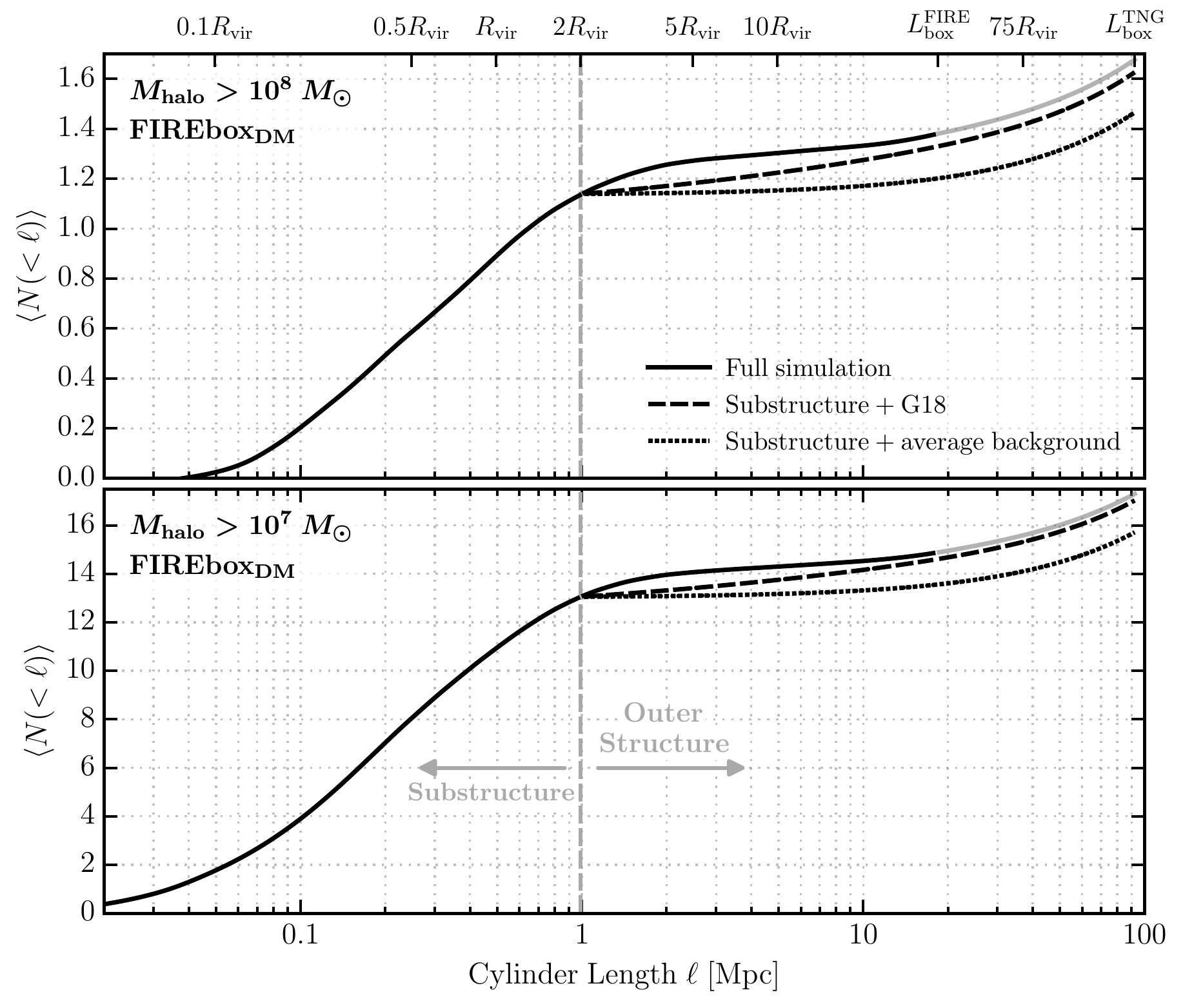}
    \caption{
    Similar to Fig.~\ref{fig:5}, but now for \fbxdm{} halos with $M_{\rm halo} > 10^{8}\ M_{\odot}$ (top) and  $M_{\rm halo} > 10^{7}\ M_{\odot}$ (bottom). Both curves (top and bottom panels) incorporating the G18 model use $\delta_{\rm los} =1$. The typical LOS passing through a lensing host encounters significantly more perturbers above a given mass as the perturber mass threshold is lowered. 
    }
    \label{fig:6}
\end{figure*}

The dashed and dotted lines in Fig.~\ref{fig:5} compares our simulation results to alternative ways of estimating cumulative counts beyond the host halo virial radius. The first assumes that halo counts are equal to the universal average \citep[estimated via][]{sheth2002excursion} for all radii beyond the virial radius of the lens (dotted lines) and the second uses the estimate from G18, which models local clustering as in Eq.~\ref{eq:analytical.los}. We see that if we assume the average background is achieved for radii beyond the virial radius, the cumulative count at 100 Mpc is underpredicted by  $\sim 20\%$ in \tng{} and $\sim 15\%$ in \tngdm. Differences between the simulations and the G18 estimate are not as large, with $\approx 5\%$ offsets in \tngdm{} and \tng{} at 100 Mpc. Note that when using the G18 formula, we set $\delta_{\rm los} =1$ for \tngdm{} and $\delta_{\rm los} =0.879$ for \tng.  The latter value is below unity because the average differential count is slightly below the \citet{sheth2002excursion} estimate at large $\ell$ (see the magenta vs. dotted lines at > 40 Mpc in Fig.~\ref{fig:4}). This factor is also introduced for the average background interpolation. 
We provide a more thorough discussion on how the contribution of the clustering component to quantifying the LOS structure of \tng{} in later in this section, while \tngdm{} results are discussed in Appendix~\ref{sec:dmo.discussion}. 

Fig.~\ref{fig:6} displays the mean cumulative counts for \fbxdm{} subhalos of $M_{\rm halo} > 10^{8}\ M_{\odot}$ and $M_{\rm halo} > 10^{7}\ M_{\odot}$ in the top and bottom panel, respectively. The likelihood of finding a halo in a single projected cylinder increases substantially at these lower masses compared to the $10^{9}\ M_{\odot}$. Specifically, we expect to see, on average, more than one small halo per LOS for \mEight{} and more than 15 for \mSeven.One caveat here is that these simulations do not include the destructive effects of a central galaxy.  We would expect substructure to be depleted within $\ell \approx 1$~Mpc  by approximately $40\%$ if the destruction mirrors that seen in Fig. \ref{fig:5} for $M_{\rm halo}>10^{9}\ M_{\odot}$ halos.\footnote{Recently, \protect\cite{kelley2019elvis} presented high-resolution zoom simulations for MW-mass DMO halos while accounting for the central galaxy and found that the depletion of substructure is about roughly the same factor for halos down to $10^{7}\ M_{\odot}$ at $z=0$. While we are drawing possible conclusions from MW-mass halos, this scaling could translate to our lens-mass halos.}

The line styles in Fig. \ref{fig:6} mimic those in Fig. \ref{fig:5}, with solid lines representing the full simulation results. The line is extrapolated beyond the edge of \fbxdm{}  (gray solid line) by assuming it follows the G18 estimate starting at $L_{\rm box}^{\rm FIRE}$ with $\delta_{\rm los} =1$ for both mass cuts. We have also modeled the entire solid line using Eq.~\eqref{eq:analytical.los.corrected} instead of the simulation results to test whether the relatively small box of \fbxdm{} suppresses large-scale modes, thereby affecting the number of halos found from $\ell = 5R_{\rm vir}$ out to \lfbx. Doing so, we find very little difference in the final results. The assumption of average background to the simulation results out to $L_{\rm box}^{\rm TNG}$ results differ to about $15\%$ for \mEight{} while $10\%$ for \mSeven. 

\begin{figure}
    \centering
    \includegraphics[width=\columnwidth]{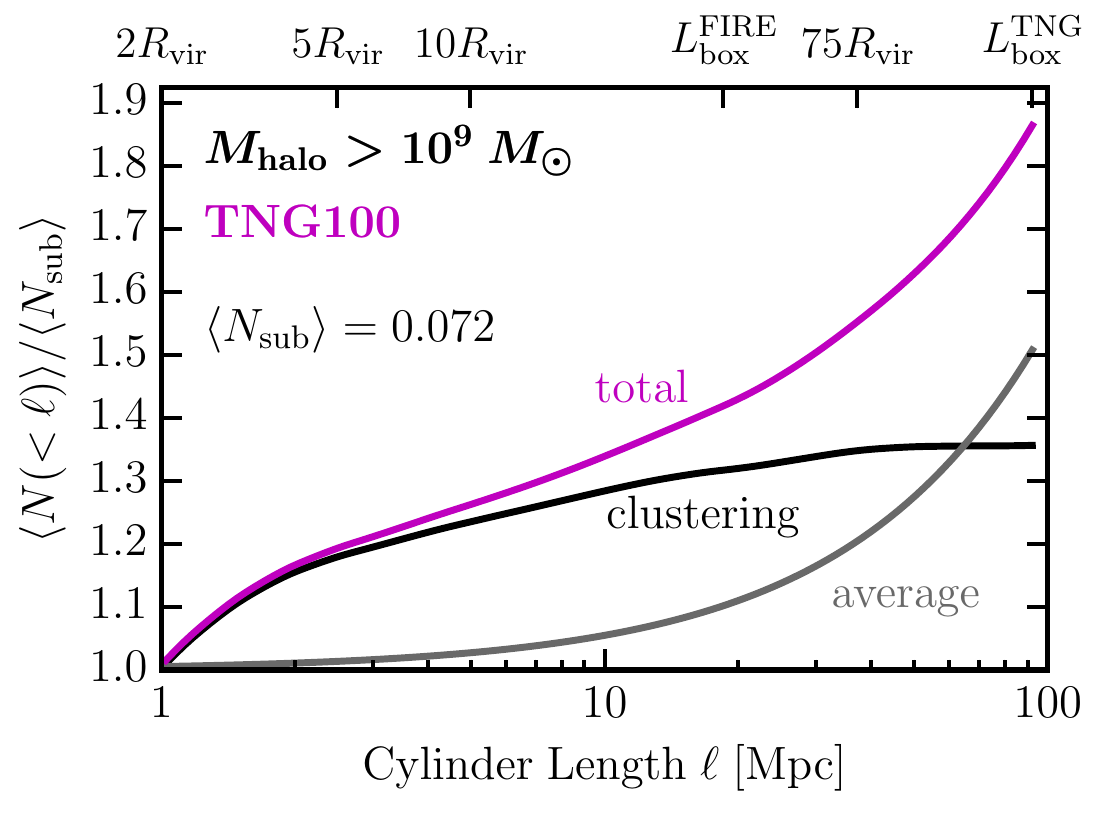}
    \caption{---
        {\bf \em Boost in counts from local clustering}: The mean count of LOS halos more massive than \mNine{} within lens-centerd cylinders from \tng{} relative to the mean count within the virial radius $\langle N_{\rm sub} \rangle \equiv \langle N(\ell< 2\,R_{\rm vir})\rangle$ is presented here.   The magenta line shows the mean total count from the simulation (labeled ``total''), while the gray line shows counts in cylinders that assume the density of halos matches the mean background (dubbed ``average'') for $\ell > 2\,R_{\rm vir}$. The black line shows the difference between the ``total" and ``average" contributions; this is the component that is attributable to local halo clustering (dubbed ``clustering'').
        We see that the local clustering effect provides a boost of $\sim 35\%$ compared to the subhalo count alone and that this contribution dominates the ``average" contribution out to ${\sim}65\,{\rm Mpc} \approx 130\,R_{\rm vir}$.
    }
    \label{fig:7}
\end{figure}

It is important to quantify the contribution of correlated clustering to counts for $\ell > 2\,R_{\rm vir}$, i.e., when subtracting off the contribution of subhalos to $N(<\ell)$, what is the fractional contribution of the population of clustered halos compared to the average halo population at $\ell > 2\,R_{\rm vir}$? This is explicitly address in Fig.~\ref{fig:7}, which shows the total average cumulative count of \mNine{} halos in \tng{} for cylinder lengths ranging from the edge of the virial radius to the edge of the box. 
We plot the average count $\langle N(< \ell) \rangle$ in units of the average cumulative count from subhalos $\langle N_{\rm sub} \rangle \equiv \langle N(\ell< 2 R_{\rm vir})\rangle$, where $\langle N_{\rm sub} \rangle = 0.072$. The full count from the simulation (labeled as ``total'') is shown by the magenta curve while cylinders that are assumed to have the average background outside of the virial radius (labeled ``average'') are plotted as the light gray curve.  To quantify the excess clustering associated beyond substructure, we take the difference between the magenta and gray curves. This results in the black curve (``clustering''). The excess clustered contribution asymptotes to $\sim 1.35\langle N_{\rm sub} \rangle$  at $\ell\approx 70\ \mathrm{Mpc}\ ({\sim}75\, R_{\rm vir})$. This means that local clustering boosts the expected signal by $\sim 35\%$ compared to what we would expect from subhalos alone.  We show similar results for lower mass halos in other simulations in Appendix~\ref{sec:dmo.discussion}. Broadly speaking, the boost from local clustering is smaller {\em relatively} in DMO simulations that do not have enhanced subhalo destruction from the central galaxy. Note that the clustering contribution is larger than the average contribution out to $\ell \approx 65\,{\rm Mpc}$, or equivalently, ${\sim}130\,R_{\rm vir}$.

\subsection{Structure along principal axes}
\label{sec:principal}
CDM halos have significant triaxiality \citep[e.g.][]{frenk1988formation,dubinski199structure,warren1992shapes,cole1996structure,jing2002triaxial,bailin2005shapes,kasun2005shapes,paz2006shapes,allgood2006shape,bett2007shape,munoz2011cdm,despali2014triaxial,vega2017shape,lau2020triaxial}, which could also mean that subhalos are found preferentially along the host's major (densest) axis \citep[e.g.][]{zentner2005anis}. This effect can qualitatively impact  our analysis of lens-centered projections, especially for substructure lensing. Furthermore, it is likely that galaxy lenses are biased to be oriented with the LOS coinciding with the host halo's major axis \citep{Mandelbaum2009,Osato2018}, as this configuration produces a larger surface mass density for a fixed overall mass distribution. In order to explore the potential magnitude of this effect, we calculate dark matter halo shapes from the lens targets using the shape inertia tensor as discussed in \cite{allgood2006shape}. We include all dark matter particles within a shell between $10-20\%$ of $R_{\rm vir}$ as a conservative approach for our sample of halos. Using a shell rather than the enclosed mass minimizes the influence of particles with radii smaller than the numerical convergence scale. The resulting eigenvalues of the shape tensor are proportional to the square root of the principal axes of the dark matter distribution. We then re-do the analysis of  Section~\ref{sec:results}, now aligning the lens-centered projections along each principal axis of the lens-mass halo.

\begin{figure}
    \centering
    \includegraphics[width=\columnwidth]{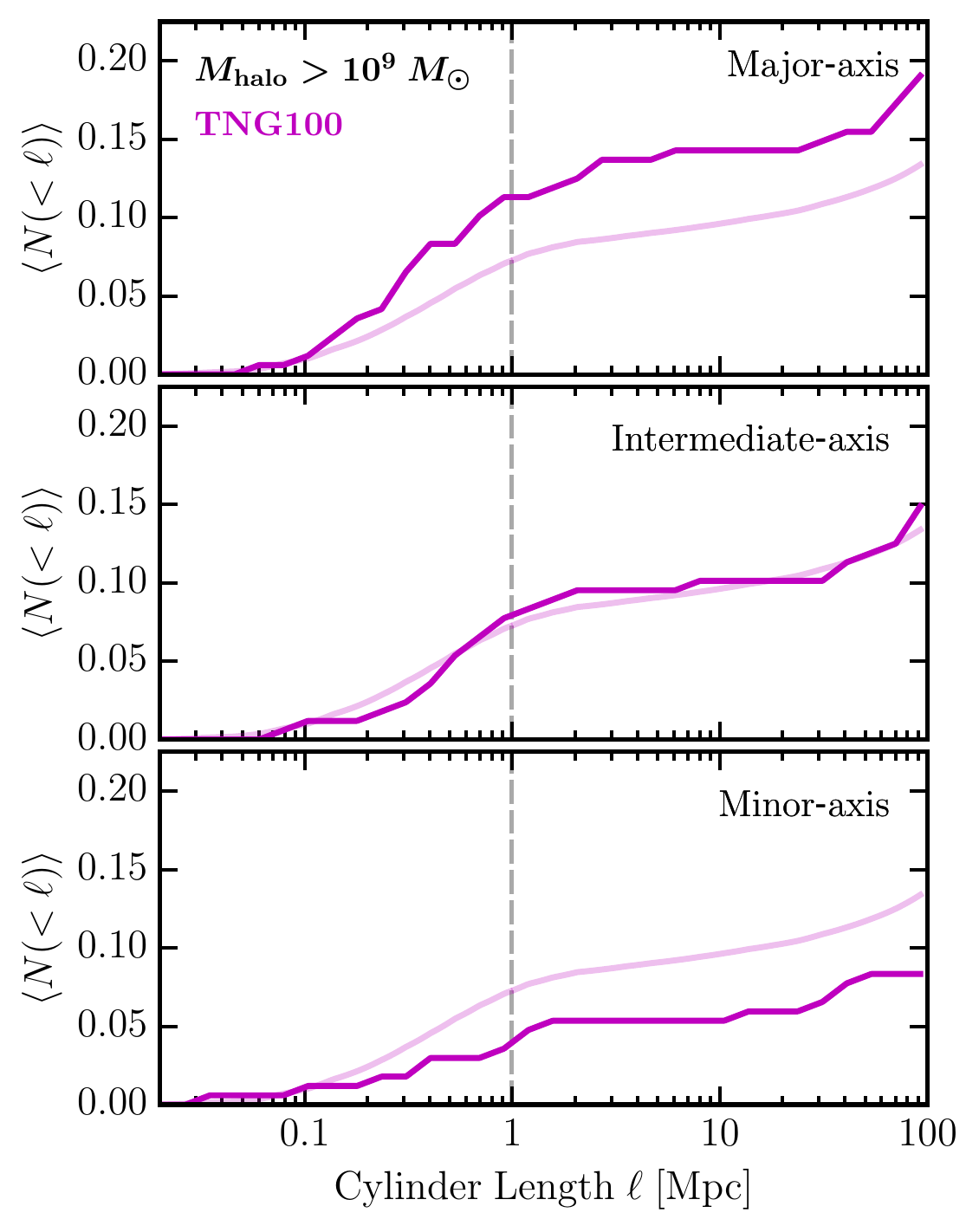}
    \caption{---
        {\bf \em Projections along the principal axes}: 
        Similar presentation as Fig.~\ref{fig:4}, but now filled solid lines show the counts for \tng{} halos $M_{\rm halo} > 10^{9}\ M_{\odot}$ along the major, intermediate, and minor axis in the top, center, and bottom panels, respectively. The transparent lines give the mean counts of the substructure with local clustering counts of \tng{} presented previously in Fig.~\ref{fig:5}. Sight-lines oriented along the major (minor) axis of the halo result in a non-trivial increase (decrease) in the number of perturbers encountered along the sight line.
    }
    \label{fig:8}
\end{figure}

Fig.~\ref{fig:8} plots the results for the average (mean) counts along each principal axis for halos more massive than \mNine{} in \tng{} (for comparison with \tngdm, see Fig.~\ref{fig:A2}). The top, middle, and bottom panels depict the average counts along the major axis, intermediate axis, and minor axis, respectively, with cylinders of radius $\mathcal{R} = 10\ \rm kpc$. For comparison, the faded solid line in each panel shows the mean counts presented previously in Fig.~\ref{fig:5}. We see that the major axis sight-line results in measurably more halos, on average, than do other orientations. The boost along the major axis is $\sim 30\%$ compared to the random LOS, with essentially the entire contribution coming from subhalos (and backsplash halos). Along the minor axis, on the other hand, average counts are significantly reduced. Projections along the intermediate-axis are comparable to the average counts shown in Fig.~\ref{fig:4}. It is clear that lens-centered projections along the major or minor axis can non-trivially boost or decrease the lensing signals by a factor of $\sim 2$. This factor is also acquired for analog simulation neglecting the presence of the central galaxy and baryons (see Fig.~\ref{fig:A2}).

\section{Summary}
\label{sec:conclusion}
Using the set of \tng{} and DMO \fbxdm{} cosmological simulations, we quantify the effect of local clustering on gravitational lensing searches for low-mass dark matter halos. We specifically focus on lens-mass halos of mass $M_{\rm vir} \simeq 10^{13}\ M_{\odot}$ at $z=0.2$ as prime targets for future lensing surveys and explore counts of halos down to $M_{\rm halo} = 10^{7-9}\ M_{\odot}$. 

Our primary result is that local clustering can boost the expected LOS perturber halo counts significantly compared subhalos alone.  The signal exceeds that expected for an average background projection to distances in excess of $\pm 10$ Mpc from the lens host (Fig. \ref{fig:4}), with a significant excess within $2-5\, R_{\rm vir}$. We provide an analytic expression for this contribution (Eq. \ref{eq:analytical.los.corrected}), which we hope will be useful in full lensing interpretation studies.

Using full-physics \tng{} simulations (which resolve halos down to \mNine), we find that the central galaxy in lens-mass hosts depletes subhalos by $\sim 70 \%$ compared to dark-matter-only simulations (see Fig. \ref{fig:5}). This result agrees with previous studies \citep{despali2017impact,graus2018smooth}. From \tng{}, the excess local clustering outside of the virial radius gives an expected count that is $\sim 35\%$ higher than would be expected from subhalos alone (Fig. \ref{fig:7}). 

Local contributions to perturber counts are also affected by halo orientation.  The above results assume a random lens-orientation with respect to the observer, but if there is a bias for lenses to be oriented along the principal axis
\citep[e.g.][]{dietrich2014orientation,groener2014shape}, then the expected local count may be enhanced. Our initial exploration of this issue indicates that local projected counts are $\sim 50\%$ higher when the target halo is oriented along the major axis compared to a random orientation (Fig. \ref{fig:8}). This result, in turn, has implications for derived constraints on the mass spectrum of perturbers and accompanying constraints on dark matter particle properties.  

The above results will be particularly important for low-redshift lenses ($z_l \sim 0.2$), such as those in the SLACS sample \citep{vegetti2014inference}.
For such lenses, \citet{despali2018modelling} find that subhalos should contribute $\sim 30-50\%$ of the total perturber signal {\em relative} to LOS halos that neglect local clustering. With clustering included, the local (subhalo + clustering) contribution may well be comparable to $\sim 67.5\%$ of the total LOS contribution that neglect clustering (or $\sim 40\%$ of the total contribution) for some lenses, especially those with lower-redshift sources ($z_s \sim 0.6$). 
Taking into account any biases in lens-halo orientation is also important for these lenses, as an additional $\sim 50\%$ boost in counts from local clustering could significantly affect interpretations. 

\section*{Acknowledgements}
This article was worked to completion during the COVID-19 lock-down and would not have been possible without the labors of our essential workers.
We thank Ran Li for helpful comments in improving the early version of this article.
We are thankful to Quinn Minor and Manoj Kaplinghat for helpful discussions. The authors thank the Illustris collaboration for facilitating the IllustrisTNG simulations for public access and the FIRE collaboration for the use of the DMO FIREbox simulation for our analysis.
AL and JSB was supported by the National Science Foundation (NSF) grant AST-1910965. AL was partially supported by NASA grant 80NSSSC20K1469.
MBK acknowledges support from NSF CAREER award AST-1752913, NSF grant AST-1910346, NASA grant NNX17AG29G, and HST-AR-15006, HST-GO-14191, HST-GO-15658, HST-GO-15901, and HST-GO-15902 from the Space Telescope Science Institute, which is operated by AURA, Inc., under NASA contract NAS5-26555. 
RF acknowledges financial support from the Swiss National Science Foundation (grant no 157591 and 194814).
The analysis in this manuscript made extensive use of the python packages {\footnotesize COLOSSUS} \citep{diemer2018colossus}, {\footnotesize NumPy} \citep{van2011numpy}, {\footnotesize SciPy} \citep{oliphant2007python}, and {\footnotesize Matplotlib} \citep{hunter2007matplotlib}; We are thankful to the developers of these tools. This research has made all intensive use of NASA's Astrophysics Data System (\url{http://ui.adsabs.harvard.edu/}) and the arXiv eprint service (\url{http://arxiv.org}).

\section*{Data Availability}
The data supporting the plots within this article are available on reasonable request to the corresponding author. A public version of the {\footnotesize Arepo} code is available at \url{https://arepo-code.org/about-arepo}. The IllustrisTNG data, including simulation snapshots and derived data products are publicly available at \url{https://www.tng-project.org/}. A public version of the {\footnotesize GIZMO} code is available at \url{http://www.tapir.caltech.edu/~phopkins/Site/GIZMO.html}. Additional data from the FIRE project, including simulation snapshots, initial conditions, and derived data products, are available at \url{https://fire.northwestern.edu/data/}.

\bibliography{references}

\begin{thebibliography}{}
\makeatletter
\relax
\def\mn@urlcharsother{\let\do\@makeother \do\$\do\&\do\#\do\^\do\_\do\%\do\~}
\def\mn@doi{\begingroup\mn@urlcharsother \@ifnextchar [ {\mn@doi@}
  {\mn@doi@[]}}
\def\mn@doi@[#1]#2{\def\@tempa{#1}\ifx\@tempa\@empty \href
  {http://dx.doi.org/#2} {doi:#2}\else \href {http://dx.doi.org/#2} {#1}\fi
  \endgroup}
\def\mn@eprint#1#2{\mn@eprint@#1:#2::\@nil}
\def\mn@eprint@arXiv#1{\href {http://arxiv.org/abs/#1} {{\tt arXiv:#1}}}
\def\mn@eprint@dblp#1{\href {http://dblp.uni-trier.de/rec/bibtex/#1.xml}
  {dblp:#1}}
\def\mn@eprint@#1:#2:#3:#4\@nil{\def\@tempa {#1}\def\@tempb {#2}\def\@tempc
  {#3}\ifx \@tempc \@empty \let \@tempc \@tempb \let \@tempb \@tempa \fi \ifx
  \@tempb \@empty \def\@tempb {arXiv}\fi \@ifundefined
  {mn@eprint@\@tempb}{\@tempb:\@tempc}{\expandafter \expandafter \csname
  mn@eprint@\@tempb\endcsname \expandafter{\@tempc}}}

\bibitem[\protect\citeauthoryear{{Allgood}, {Flores}, {Primack}, {Kravtsov},
  {Wechsler}, {Faltenbacher}  \& {Bullock}}{{Allgood}
  et~al.}{2006}]{allgood2006shape}
{Allgood} B.,  {Flores} R.~A.,  {Primack} J.~R.,  {Kravtsov} A.~V.,  {Wechsler}
  R.~H.,  {Faltenbacher} A.,   {Bullock} J.~S.,  2006, \mn@doi [\mnras]
  {10.1111/j.1365-2966.2006.10094.x}, \href
  {https://ui.adsabs.harvard.edu/abs/2006MNRAS.367.1781A} {367, 1781}

\bibitem[\protect\citeauthoryear{{Bailin} \& {Steinmetz}}{{Bailin} \&
  {Steinmetz}}{2005}]{bailin2005shapes}
{Bailin} J.,  {Steinmetz} M.,  2005, \mn@doi [\apj] {10.1086/430397}, \href
  {https://ui.adsabs.harvard.edu/abs/2005ApJ...627..647B} {627, 647}

\bibitem[\protect\citeauthoryear{{Banik}, {Bovy}, {Bertone}, {Erkal}  \& {de
  Boer}}{{Banik} et~al.}{2019}]{banik2019novel}
{Banik} N.,  {Bovy} J.,  {Bertone} G.,  {Erkal} D.,   {de Boer} T.~J.~L.,
  2019, arXiv e-prints, \href
  {https://ui.adsabs.harvard.edu/abs/2019arXiv191102663B} {p. arXiv:1911.02663}

\bibitem[\protect\citeauthoryear{{Behroozi}, {Wechsler}  \& {Wu}}{{Behroozi}
  et~al.}{2013}]{behroozi2012rockstar}
{Behroozi} P.~S.,  {Wechsler} R.~H.,   {Wu} H.-Y.,  2013, \mn@doi [\apj]
  {10.1088/0004-637X/762/2/109}, \href
  {https://ui.adsabs.harvard.edu/abs/2013ApJ...762..109B} {762, 109}

\bibitem[\protect\citeauthoryear{{Bertone} \& {Tait}}{{Bertone} \&
  {Tait}}{2018}]{bertone2018era}
{Bertone} G.,  {Tait} T. M.~P.,  2018, \mn@doi [\nat]
  {10.1038/s41586-018-0542-z}, \href
  {https://ui.adsabs.harvard.edu/abs/2018Natur.562...51B} {562, 51}

\bibitem[\protect\citeauthoryear{{Bett}, {Eke}, {Frenk}, {Jenkins}, {Helly}  \&
  {Navarro}}{{Bett} et~al.}{2007}]{bett2007shape}
{Bett} P.,  {Eke} V.,  {Frenk} C.~S.,  {Jenkins} A.,  {Helly} J.,   {Navarro}
  J.,  2007, \mn@doi [\mnras] {10.1111/j.1365-2966.2007.11432.x}, \href
  {https://ui.adsabs.harvard.edu/abs/2007MNRAS.376..215B} {376, 215}

\bibitem[\protect\citeauthoryear{{Bode}, {Ostriker}  \& {Turok}}{{Bode}
  et~al.}{2001}]{bode2001wdm}
{Bode} P.,  {Ostriker} J.~P.,   {Turok} N.,  2001, \mn@doi [\apj]
  {10.1086/321541}, \href
  {https://ui.adsabs.harvard.edu/abs/2001ApJ...556...93B} {556, 93}

\bibitem[\protect\citeauthoryear{{Bolton}, {Burles}, {Koopmans}, {Treu},
  {Gavazzi}, {Moustakas}, {Wayth}  \& {Schlegel}}{{Bolton}
  et~al.}{2008}]{bolton2008sloan}
{Bolton} A.~S.,  {Burles} S.,  {Koopmans} L. V.~E.,  {Treu} T.,  {Gavazzi} R.,
  {Moustakas} L.~A.,  {Wayth} R.,   {Schlegel} D.~J.,  2008, \mn@doi [\apj]
  {10.1086/589327}, \href
  {https://ui.adsabs.harvard.edu/abs/2008ApJ...682..964B} {682, 964}

\bibitem[\protect\citeauthoryear{{Bond}, {Kofman}  \& {Pogosyan}}{{Bond}
  et~al.}{1996}]{bond1996filaments}
{Bond} J.~R.,  {Kofman} L.,   {Pogosyan} D.,  1996, \mn@doi [\nat]
  {10.1038/380603a0}, \href
  {https://ui.adsabs.harvard.edu/abs/1996Natur.380..603B} {380, 603}

\bibitem[\protect\citeauthoryear{{Bozek}, {Boylan-Kolchin}, {Horiuchi},
  {Garrison-Kimmel}, {Abazajian}  \& {Bullock}}{{Bozek}
  et~al.}{2016}]{bozek2016resonant}
{Bozek} B.,  {Boylan-Kolchin} M.,  {Horiuchi} S.,  {Garrison-Kimmel} S.,
  {Abazajian} K.,   {Bullock} J.~S.,  2016, \mn@doi [\mnras]
  {10.1093/mnras/stw688}, \href
  {https://ui.adsabs.harvard.edu/abs/2016MNRAS.459.1489B} {459, 1489}

\bibitem[\protect\citeauthoryear{{Brada{\v{c}}}, {Schneider}, {Steinmetz},
  {Lombardi}, {King}  \& {Porcas}}{{Brada{\v{c}}}
  et~al.}{2002}]{bradac2002mass}
{Brada{\v{c}}} M.,  {Schneider} P.,  {Steinmetz} M.,  {Lombardi} M.,  {King}
  L.~J.,   {Porcas} R.,  2002, \mn@doi [\aap] {10.1051/0004-6361:20020559},
  \href {https://ui.adsabs.harvard.edu/abs/2002A&A...388..373B} {388, 373}

\bibitem[\protect\citeauthoryear{{Brooks} \& {Zolotov}}{{Brooks} \&
  {Zolotov}}{2014}]{brooks2014baryons}
{Brooks} A.~M.,  {Zolotov} A.,  2014, \mn@doi [\apj]
  {10.1088/0004-637X/786/2/87}, \href
  {https://ui.adsabs.harvard.edu/abs/2014ApJ...786...87B} {786, 87}

\bibitem[\protect\citeauthoryear{{Bryan} \& {Norman}}{{Bryan} \&
  {Norman}}{1998}]{bryan1998statistical}
{Bryan} G.~L.,  {Norman} M.~L.,  1998, \mn@doi [\apj] {10.1086/305262}, \href
  {https://ui.adsabs.harvard.edu/abs/1998ApJ...495...80B} {495, 80}

\bibitem[\protect\citeauthoryear{{Bullock} \& {Boylan-Kolchin}}{{Bullock} \&
  {Boylan-Kolchin}}{2017}]{bullock2017small}
{Bullock} J.~S.,  {Boylan-Kolchin} M.,  2017, \mn@doi [\araa]
  {10.1146/annurev-astro-091916-055313}, \href
  {https://ui.adsabs.harvard.edu/abs/2017ARA&A..55..343B} {55, 343}

\bibitem[\protect\citeauthoryear{{Bullock}, {Kravtsov}  \&
  {Weinberg}}{{Bullock} et~al.}{2000}]{bullock2000}
{Bullock} J.~S.,  {Kravtsov} A.~V.,   {Weinberg} D.~H.,  2000, \mn@doi [\apj]
  {10.1086/309279}, \href
  {https://ui.adsabs.harvard.edu/abs/2000ApJ...539..517B} {539, 517}

\bibitem[\protect\citeauthoryear{{Carlberg}}{{Carlberg}}{2009}]{calberg2009streams}
{Carlberg} R.~G.,  2009, \mn@doi [\apjl] {10.1088/0004-637X/705/2/L223}, \href
  {https://ui.adsabs.harvard.edu/abs/2009ApJ...705L.223C} {705, L223}

\bibitem[\protect\citeauthoryear{{Cole} \& {Lacey}}{{Cole} \&
  {Lacey}}{1996}]{cole1996structure}
{Cole} S.,  {Lacey} C.,  1996, \mn@doi [\mnras] {10.1093/mnras/281.2.716},
  \href {https://ui.adsabs.harvard.edu/abs/1996MNRAS.281..716C} {281, 716}

\bibitem[\protect\citeauthoryear{{Collett}}{{Collett}}{2015}]{collett2015forthcoming}
{Collett} T.~E.,  2015, \mn@doi [\apj] {10.1088/0004-637X/811/1/20}, \href
  {https://ui.adsabs.harvard.edu/abs/2015ApJ...811...20C} {811, 20}

\bibitem[\protect\citeauthoryear{{D'Aloisio}, {Natarajan}  \&
  {Shapiro}}{{D'Aloisio} et~al.}{2014}]{daloisio2014effect}
{D'Aloisio} A.,  {Natarajan} P.,   {Shapiro} P.~R.,  2014, \mn@doi [\mnras]
  {10.1093/mnras/stu1931}, \href
  {https://ui.adsabs.harvard.edu/abs/2014MNRAS.445.3581D} {445, 3581}

\bibitem[\protect\citeauthoryear{{Dalal} \& {Kochanek}}{{Dalal} \&
  {Kochanek}}{2002}]{dalal2002direct}
{Dalal} N.,  {Kochanek} C.~S.,  2002, \mn@doi [\apj] {10.1086/340303}, \href
  {https://ui.adsabs.harvard.edu/abs/2002ApJ...572...25D} {572, 25}

\bibitem[\protect\citeauthoryear{{Davis}, {Efstathiou}, {Frenk}  \&
  {White}}{{Davis} et~al.}{1985}]{davis1985cdm}
{Davis} M.,  {Efstathiou} G.,  {Frenk} C.~S.,   {White} S.~D.~M.,  1985,
  \mn@doi [\apj] {10.1086/163168}, \href
  {https://ui.adsabs.harvard.edu/abs/1985ApJ...292..371D} {292, 371}

\bibitem[\protect\citeauthoryear{{Despali} \& {Vegetti}}{{Despali} \&
  {Vegetti}}{2017}]{despali2017impact}
{Despali} G.,  {Vegetti} S.,  2017, \mn@doi [\mnras] {10.1093/mnras/stx966},
  \href {https://ui.adsabs.harvard.edu/abs/2017MNRAS.469.1997D} {469, 1997}

\bibitem[\protect\citeauthoryear{{Despali}, {Giocoli}  \& {Tormen}}{{Despali}
  et~al.}{2014}]{despali2014triaxial}
{Despali} G.,  {Giocoli} C.,   {Tormen} G.,  2014, \mn@doi [\mnras]
  {10.1093/mnras/stu1393}, \href
  {https://ui.adsabs.harvard.edu/abs/2014MNRAS.443.3208D} {443, 3208}

\bibitem[\protect\citeauthoryear{{Despali}, {Vegetti}, {White}, {Giocoli}  \&
  {van den Bosch}}{{Despali} et~al.}{2018}]{despali2018modelling}
{Despali} G.,  {Vegetti} S.,  {White} S. D.~M.,  {Giocoli} C.,   {van den
  Bosch} F.~C.,  2018, \mn@doi [\mnras] {10.1093/mnras/sty159}, \href
  {https://ui.adsabs.harvard.edu/abs/2018MNRAS.475.5424D} {475, 5424}

\bibitem[\protect\citeauthoryear{{Diemer}}{{Diemer}}{2018}]{diemer2018colossus}
{Diemer} B.,  2018, \mn@doi [\apjs] {10.3847/1538-4365/aaee8c}, \href
  {https://ui.adsabs.harvard.edu/abs/2018ApJS..239...35D} {239, 35}

\bibitem[\protect\citeauthoryear{{Dietrich} et~al.,}{{Dietrich}
  et~al.}{2014}]{dietrich2014orientation}
{Dietrich} J.~P.,  et~al., 2014, \mn@doi [\mnras] {10.1093/mnras/stu1282},
  \href {https://ui.adsabs.harvard.edu/abs/2014MNRAS.443.1713D} {443, 1713}

\bibitem[\protect\citeauthoryear{{Dolag}, {Komatsu}  \& {Sunyaev}}{{Dolag}
  et~al.}{2016}]{dolag2016pathfinder}
{Dolag} K.,  {Komatsu} E.,   {Sunyaev} R.,  2016, \mn@doi [\mnras]
  {10.1093/mnras/stw2035}, \href
  {https://ui.adsabs.harvard.edu/abs/2016MNRAS.463.1797D} {463, 1797}

\bibitem[\protect\citeauthoryear{{Dubinski} \& {Carlberg}}{{Dubinski} \&
  {Carlberg}}{1991}]{dubinski199structure}
{Dubinski} J.,  {Carlberg} R.~G.,  1991, \mn@doi [\apj] {10.1086/170451}, \href
  {https://ui.adsabs.harvard.edu/abs/1991ApJ...378..496D} {378, 496}

\bibitem[\protect\citeauthoryear{{Dubois}, {Peirani}, {Pichon}, {Devriendt},
  {Gavazzi}, {Welker}  \& {Volonteri}}{{Dubois}
  et~al.}{2016}]{dubois2016horizon}
{Dubois} Y.,  {Peirani} S.,  {Pichon} C.,  {Devriendt} J.,  {Gavazzi} R.,
  {Welker} C.,   {Volonteri} M.,  2016, \mn@doi [\mnras]
  {10.1093/mnras/stw2265}, \href
  {https://ui.adsabs.harvard.edu/abs/2016MNRAS.463.3948D} {463, 3948}

\bibitem[\protect\citeauthoryear{{Efstathiou}}{{Efstathiou}}{1992}]{efstathiou1992}
{Efstathiou} G.,  1992, \mn@doi [\mnras] {10.1093/mnras/256.1.43P}, \href
  {https://ui.adsabs.harvard.edu/abs/1992MNRAS.256P..43E} {256, 43P}

\bibitem[\protect\citeauthoryear{{Feldmann} \& {Spolyar}}{{Feldmann} \&
  {Spolyar}}{2015}]{feldmann2015gaia}
{Feldmann} R.,  {Spolyar} D.,  2015, \mn@doi [\mnras] {10.1093/mnras/stu2147},
  \href {https://ui.adsabs.harvard.edu/abs/2015MNRAS.446.1000F} {446, 1000}

\bibitem[\protect\citeauthoryear{{Frenk}, {White}, {Davis}  \&
  {Efstathiou}}{{Frenk} et~al.}{1988}]{frenk1988formation}
{Frenk} C.~S.,  {White} S. D.~M.,  {Davis} M.,   {Efstathiou} G.,  1988,
  \mn@doi [\apj] {10.1086/166213}, \href
  {https://ui.adsabs.harvard.edu/abs/1988ApJ...327..507F} {327, 507}

\bibitem[\protect\citeauthoryear{{Garrison-Kimmel}, {Boylan-Kolchin}, {Bullock}
   \& {Lee}}{{Garrison-Kimmel} et~al.}{2014}]{sgk2014elvis}
{Garrison-Kimmel} S.,  {Boylan-Kolchin} M.,  {Bullock} J.~S.,   {Lee} K.,
  2014, \mn@doi [\mnras] {10.1093/mnras/stt2377}, \href
  {https://ui.adsabs.harvard.edu/abs/2014MNRAS.438.2578G} {438, 2578}

\bibitem[\protect\citeauthoryear{{Garrison-Kimmel} et~al.,}{{Garrison-Kimmel}
  et~al.}{2017}]{sgk2017lumpy}
{Garrison-Kimmel} S.,  et~al., 2017, \mn@doi [\mnras] {10.1093/mnras/stx1710},
  \href {https://ui.adsabs.harvard.edu/abs/2017MNRAS.471.1709G} {471, 1709}

\bibitem[\protect\citeauthoryear{{Geller} \& {Huchra}}{{Geller} \&
  {Huchra}}{1989}]{geller1989mapping}
{Geller} M.~J.,  {Huchra} J.~P.,  1989, \mn@doi [Science]
  {10.1126/science.246.4932.897}, \href
  {https://ui.adsabs.harvard.edu/abs/1989Sci...246..897G} {246, 897}

\bibitem[\protect\citeauthoryear{{Genel} et~al.,}{{Genel}
  et~al.}{2014}]{genel2014illustris}
{Genel} S.,  et~al., 2014, \mn@doi [\mnras] {10.1093/mnras/stu1654}, \href
  {https://ui.adsabs.harvard.edu/abs/2014MNRAS.445..175G} {445, 175}

\bibitem[\protect\citeauthoryear{{Gill}, {Knebe}  \& {Gibson}}{{Gill}
  et~al.}{2005}]{gill2005outskirts}
{Gill} S. P.~D.,  {Knebe} A.,   {Gibson} B.~K.,  2005, \mn@doi [\mnras]
  {10.1111/j.1365-2966.2004.08562.x}, \href
  {https://ui.adsabs.harvard.edu/abs/2005MNRAS.356.1327G} {356, 1327}

\bibitem[\protect\citeauthoryear{{Gilman}, {Agnello}, {Treu}, {Keeton}  \&
  {Nierenberg}}{{Gilman} et~al.}{2017}]{gilman2017strong}
{Gilman} D.,  {Agnello} A.,  {Treu} T.,  {Keeton} C.~R.,   {Nierenberg} A.~M.,
  2017, \mn@doi [\mnras] {10.1093/mnras/stx158}, \href
  {https://ui.adsabs.harvard.edu/abs/2017MNRAS.467.3970G} {467, 3970}

\bibitem[\protect\citeauthoryear{{Gilman}, {Birrer}, {Treu}, {Keeton}  \&
  {Nierenberg}}{{Gilman} et~al.}{2018}]{gilman2018forward}
{Gilman} D.,  {Birrer} S.,  {Treu} T.,  {Keeton} C.~R.,   {Nierenberg} A.,
  2018, \mn@doi [\mnras] {10.1093/mnras/sty2261}, \href
  {https://ui.adsabs.harvard.edu/abs/2018MNRAS.481..819G} {481, 819}

\bibitem[\protect\citeauthoryear{{Gilman}, {Birrer}, {Treu}, {Nierenberg}  \&
  {Benson}}{{Gilman} et~al.}{2019}]{gilman2019flux}
{Gilman} D.,  {Birrer} S.,  {Treu} T.,  {Nierenberg} A.,   {Benson} A.,  2019,
  \mn@doi [\mnras] {10.1093/mnras/stz1593}, \href
  {https://ui.adsabs.harvard.edu/abs/2019MNRAS.487.5721G} {487, 5721}

\bibitem[\protect\citeauthoryear{{Graus}, {Bullock}, {Boylan-Kolchin}  \&
  {Nierenberg}}{{Graus} et~al.}{2018}]{graus2018smooth}
{Graus} A.~S.,  {Bullock} J.~S.,  {Boylan-Kolchin} M.,   {Nierenberg} A.~M.,
  2018, \mn@doi [Monthly Notices of the Royal Astronomical Society]
  {10.1093/mnras/sty1924}, \href
  {https://ui.adsabs.harvard.edu/abs/2018MNRAS.480.1322G} {480, 1322}

\bibitem[\protect\citeauthoryear{{Groener} \& {Goldberg}}{{Groener} \&
  {Goldberg}}{2014}]{groener2014shape}
{Groener} A.~M.,  {Goldberg} D.~M.,  2014, \mn@doi [\apj]
  {10.1088/0004-637X/795/2/153}, \href
  {https://ui.adsabs.harvard.edu/abs/2014ApJ...795..153G} {795, 153}

\bibitem[\protect\citeauthoryear{{He}, {Li}, {Lim}, {Frenk}, {Cole}, {Peng}  \&
  {Wang}}{{He} et~al.}{2018}]{he2018globular}
{He} Q.,  {Li} R.,  {Lim} S.,  {Frenk} C.~S.,  {Cole} S.,  {Peng} E.~W.,
  {Wang} Q.,  2018, \mn@doi [\mnras] {10.1093/mnras/sty2260}, \href
  {https://ui.adsabs.harvard.edu/abs/2018MNRAS.480.5084H} {480, 5084}

\bibitem[\protect\citeauthoryear{{He} et~al.,}{{He}
  et~al.}{2020}]{he2020forward}
{He} Q.,  et~al., 2020, arXiv e-prints, \href
  {https://ui.adsabs.harvard.edu/abs/2020arXiv201013221H} {p. arXiv:2010.13221}

\bibitem[\protect\citeauthoryear{{Hezaveh} et~al.,}{{Hezaveh}
  et~al.}{2016}]{hezaveh2016detection}
{Hezaveh} Y.~D.,  et~al., 2016, \mn@doi [\apj] {10.3847/0004-637X/823/1/37},
  \href {https://ui.adsabs.harvard.edu/abs/2016ApJ...823...37H} {823, 37}

\bibitem[\protect\citeauthoryear{{Horiuchi}, {Bozek}, {Abazajian},
  {Boylan-Kolchin}, {Bullock}, {Garrison-Kimmel}  \& {Onorbe}}{{Horiuchi}
  et~al.}{2016}]{horiuchi2016properties}
{Horiuchi} S.,  {Bozek} B.,  {Abazajian} K.~N.,  {Boylan-Kolchin} M.,
  {Bullock} J.~S.,  {Garrison-Kimmel} S.,   {Onorbe} J.,  2016, \mn@doi
  [\mnras] {10.1093/mnras/stv2922}, \href
  {https://ui.adsabs.harvard.edu/abs/2016MNRAS.456.4346H} {456, 4346}

\bibitem[\protect\citeauthoryear{{Hsueh} et~al.,}{{Hsueh}
  et~al.}{2017}]{hsueh2017sharpiv}
{Hsueh} J.~W.,  et~al., 2017, \mn@doi [\mnras] {10.1093/mnras/stx1082}, \href
  {https://ui.adsabs.harvard.edu/abs/2017MNRAS.469.3713H} {469, 3713}

\bibitem[\protect\citeauthoryear{{Hsueh}, {Despali}, {Vegetti}, {Xu},
  {Fassnacht}  \& {Metcalf}}{{Hsueh} et~al.}{2018}]{hsueh2018flux}
{Hsueh} J.-W.,  {Despali} G.,  {Vegetti} S.,  {Xu} D.~a.,  {Fassnacht} C.~D.,
  {Metcalf} R.~B.,  2018, \mn@doi [\mnras] {10.1093/mnras/stx3320}, \href
  {https://ui.adsabs.harvard.edu/abs/2018MNRAS.475.2438H} {475, 2438}

\bibitem[\protect\citeauthoryear{{Hunter}}{{Hunter}}{2007}]{hunter2007matplotlib}
{Hunter} J.~D.,  2007, \mn@doi [Computing in Science and Engineering]
  {10.1109/MCSE.2007.55}, \href
  {https://ui.adsabs.harvard.edu/abs/2007CSE.....9...90H} {9, 90}

\bibitem[\protect\citeauthoryear{{Ibata}, {Lewis}, {Irwin}  \& {Quinn}}{{Ibata}
  et~al.}{2002}]{ibata2002streams}
{Ibata} R.~A.,  {Lewis} G.~F.,  {Irwin} M.~J.,   {Quinn} T.,  2002, \mn@doi
  [\mnras] {10.1046/j.1365-8711.2002.05358.x}, \href
  {https://ui.adsabs.harvard.edu/abs/2002MNRAS.332..915I} {332, 915}

\bibitem[\protect\citeauthoryear{{Inoue}}{{Inoue}}{2016}]{inoue2016origin}
{Inoue} K.~T.,  2016, \mn@doi [\mnras] {10.1093/mnras/stw1270}, \href
  {https://ui.adsabs.harvard.edu/abs/2016MNRAS.461..164I} {461, 164}

\bibitem[\protect\citeauthoryear{{Inoue} \& {Takahashi}}{{Inoue} \&
  {Takahashi}}{2012}]{inoue2012weak}
{Inoue} K.~T.,  {Takahashi} R.,  2012, \mn@doi [\mnras]
  {10.1111/j.1365-2966.2012.21915.x}, \href
  {https://ui.adsabs.harvard.edu/abs/2012MNRAS.426.2978I} {426, 2978}

\bibitem[\protect\citeauthoryear{{Inoue}, {Takahashi}, {Takahashi}  \&
  {Ishiyama}}{{Inoue} et~al.}{2015}]{inoue2015constraints}
{Inoue} K.~T.,  {Takahashi} R.,  {Takahashi} T.,   {Ishiyama} T.,  2015,
  \mn@doi [\mnras] {10.1093/mnras/stv194}, \href
  {https://ui.adsabs.harvard.edu/abs/2015MNRAS.448.2704I} {448, 2704}

\bibitem[\protect\citeauthoryear{{Jing} \& {Suto}}{{Jing} \&
  {Suto}}{2002}]{jing2002triaxial}
{Jing} Y.~P.,  {Suto} Y.,  2002, \mn@doi [\apj] {10.1086/341065}, \href
  {https://ui.adsabs.harvard.edu/abs/2002ApJ...574..538J} {574, 538}

\bibitem[\protect\citeauthoryear{{Kasun} \& {Evrard}}{{Kasun} \&
  {Evrard}}{2005}]{kasun2005shapes}
{Kasun} S.~F.,  {Evrard} A.~E.,  2005, \mn@doi [\apj] {10.1086/430811}, \href
  {https://ui.adsabs.harvard.edu/abs/2005ApJ...629..781K} {629, 781}

\bibitem[\protect\citeauthoryear{{Kelley}, {Bullock}, {Garrison-Kimmel},
  {Boylan-Kolchin}, {Pawlowski}  \& {Graus}}{{Kelley}
  et~al.}{2019}]{kelley2019elvis}
{Kelley} T.,  {Bullock} J.~S.,  {Garrison-Kimmel} S.,  {Boylan-Kolchin} M.,
  {Pawlowski} M.~S.,   {Graus} A.~S.,  2019, \mn@doi [\mnras]
  {10.1093/mnras/stz1553}, \href
  {https://ui.adsabs.harvard.edu/abs/2019MNRAS.487.4409K} {487, 4409}

\bibitem[\protect\citeauthoryear{{Khandai}, {Di Matteo}, {Croft}, {Wilkins},
  {Feng}, {Tucker}, {DeGraf}  \& {Liu}}{{Khandai}
  et~al.}{2015}]{khandai2015massiveblack}
{Khandai} N.,  {Di Matteo} T.,  {Croft} R.,  {Wilkins} S.,  {Feng} Y.,
  {Tucker} E.,  {DeGraf} C.,   {Liu} M.-S.,  2015, \mn@doi [\mnras]
  {10.1093/mnras/stv627}, \href
  {https://ui.adsabs.harvard.edu/abs/2015MNRAS.450.1349K} {450, 1349}

\bibitem[\protect\citeauthoryear{{Koopmans}}{{Koopmans}}{2005}]{koopmans2005imaging}
{Koopmans} L.~V.~E.,  2005, \mn@doi [\mnras]
  {10.1111/j.1365-2966.2005.09523.x}, \href
  {https://ui.adsabs.harvard.edu/abs/2005MNRAS.363.1136K} {363, 1136}

\bibitem[\protect\citeauthoryear{{Lau}, {Hearin}, {Nagai}  \&
  {Cappelluti}}{{Lau} et~al.}{2020}]{lau2020triaxial}
{Lau} E.~T.,  {Hearin} A.~P.,  {Nagai} D.,   {Cappelluti} N.,  2020, \mn@doi
  [\mnras] {10.1093/mnras/staa3313}, \href
  {https://ui.adsabs.harvard.edu/abs/2020MNRAS.tmp.3106L} {}

\bibitem[\protect\citeauthoryear{{Li}, {Frenk}, {Cole}, {Gao}, {Bose}  \&
  {Hellwing}}{{Li} et~al.}{2016}]{li2016constraints}
{Li} R.,  {Frenk} C.~S.,  {Cole} S.,  {Gao} L.,  {Bose} S.,   {Hellwing} W.~A.,
   2016, \mn@doi [\mnras] {10.1093/mnras/stw939}, \href
  {https://ui.adsabs.harvard.edu/abs/2016MNRAS.460..363L} {460, 363}

\bibitem[\protect\citeauthoryear{{Li}, {Frenk}, {Cole}, {Wang}  \& {Gao}}{{Li}
  et~al.}{2017}]{li2017lens}
{Li} R.,  {Frenk} C.~S.,  {Cole} S.,  {Wang} Q.,   {Gao} L.,  2017, \mn@doi
  [\mnras] {10.1093/mnras/stx554}, \href
  {https://ui.adsabs.harvard.edu/abs/2017MNRAS.468.1426L} {468, 1426}

\bibitem[\protect\citeauthoryear{{Ma} \& {Fry}}{{Ma} \&
  {Fry}}{2000}]{ma2000stabilize}
{Ma} C.-P.,  {Fry} J.~N.,  2000, \mn@doi [\apjl] {10.1086/312819}, \href
  {https://ui.adsabs.harvard.edu/abs/2000ApJ...538L.107M} {538, L107}

\bibitem[\protect\citeauthoryear{{MacLeod}, {Jones}, {Agol}  \&
  {Kochanek}}{{MacLeod} et~al.}{2013}]{macleod2013detection}
{MacLeod} C.~L.,  {Jones} R.,  {Agol} E.,   {Kochanek} C.~S.,  2013, \mn@doi
  [\apj] {10.1088/0004-637X/773/1/35}, \href
  {https://ui.adsabs.harvard.edu/abs/2013ApJ...773...35M} {773, 35}

\bibitem[\protect\citeauthoryear{{Mandelbaum}, {van de Ven}  \&
  {Keeton}}{{Mandelbaum} et~al.}{2009}]{Mandelbaum2009}
{Mandelbaum} R.,  {van de Ven} G.,   {Keeton} C.~R.,  2009, \mn@doi [\mnras]
  {10.1111/j.1365-2966.2009.15166.x}, \href
  {https://ui.adsabs.harvard.edu/abs/2009MNRAS.398..635M} {398, 635}

\bibitem[\protect\citeauthoryear{{Marinacci} et~al.,}{{Marinacci}
  et~al.}{2018}]{marinacci2018first}
{Marinacci} F.,  et~al., 2018, \mn@doi [\mnras] {10.1093/mnras/sty2206}, \href
  {https://ui.adsabs.harvard.edu/abs/2018MNRAS.480.5113M} {480, 5113}

\bibitem[\protect\citeauthoryear{{Meneghetti} et~al.,}{{Meneghetti}
  et~al.}{2020}]{meneghetti2020excess}
{Meneghetti} M.,  et~al., 2020, \mn@doi [Science] {10.1126/science.aax5164},
  \href {https://ui.adsabs.harvard.edu/abs/2020Sci...369.1347M} {369, 1347}

\bibitem[\protect\citeauthoryear{{Metcalf}}{{Metcalf}}{2005}]{metcalf2005importance}
{Metcalf} R.~B.,  2005, \mn@doi [\apj] {10.1086/431574}, \href
  {https://ui.adsabs.harvard.edu/abs/2005ApJ...629..673M} {629, 673}

\bibitem[\protect\citeauthoryear{{Metcalf} \& {Amara}}{{Metcalf} \&
  {Amara}}{2012}]{metcalf2012small}
{Metcalf} R.~B.,  {Amara} A.,  2012, \mn@doi [\mnras]
  {10.1111/j.1365-2966.2011.19982.x}, \href
  {https://ui.adsabs.harvard.edu/abs/2012MNRAS.419.3414M} {419, 3414}

\bibitem[\protect\citeauthoryear{{More}, {Diemer}  \& {Kravtsov}}{{More}
  et~al.}{2015}]{more2015splash}
{More} S.,  {Diemer} B.,   {Kravtsov} A.~V.,  2015, \mn@doi [\apj]
  {10.1088/0004-637X/810/1/36}, \href
  {https://ui.adsabs.harvard.edu/abs/2015ApJ...810...36M} {810, 36}

\bibitem[\protect\citeauthoryear{{Mu{\~n}oz-Cuartas}, {Macci{\`o}},
  {Gottl{\"o}ber}  \& {Dutton}}{{Mu{\~n}oz-Cuartas}
  et~al.}{2011}]{munoz2011cdm}
{Mu{\~n}oz-Cuartas} J.~C.,  {Macci{\`o}} A.~V.,  {Gottl{\"o}ber} S.,   {Dutton}
  A.~A.,  2011, \mn@doi [\mnras] {10.1111/j.1365-2966.2010.17704.x}, \href
  {https://ui.adsabs.harvard.edu/abs/2011MNRAS.411..584M} {411, 584}

\bibitem[\protect\citeauthoryear{{Nadler}, {Birrer}, {Gilman}, {Wechsler},
  {Du}, {Benson}, {Nierenberg}  \& {Treu}}{{Nadler}
  et~al.}{2021}]{nadler2021lensing}
{Nadler} E.~O.,  {Birrer} S.,  {Gilman} D.,  {Wechsler} R.~H.,  {Du} X.,
  {Benson} A.,  {Nierenberg} A.~M.,   {Treu} T.,  2021, arXiv e-prints, \href
  {https://ui.adsabs.harvard.edu/abs/2021arXiv210107810N} {p. arXiv:2101.07810}

\bibitem[\protect\citeauthoryear{{Naiman} et~al.,}{{Naiman}
  et~al.}{2018}]{naiman2018first}
{Naiman} J.~P.,  et~al., 2018, \mn@doi [\mnras] {10.1093/mnras/sty618}, \href
  {https://ui.adsabs.harvard.edu/abs/2018MNRAS.477.1206N} {477, 1206}

\bibitem[\protect\citeauthoryear{{Nelson} et~al.,}{{Nelson}
  et~al.}{2018}]{nelson2017first}
{Nelson} D.,  et~al., 2018, \mn@doi [\mnras] {10.1093/mnras/stx3040}, \href
  {https://ui.adsabs.harvard.edu/abs/2018MNRAS.475..624N} {475, 624}

\bibitem[\protect\citeauthoryear{{Nierenberg} et~al.,}{{Nierenberg}
  et~al.}{2017}]{nierenberg2017probing}
{Nierenberg} A.~M.,  et~al., 2017, \mn@doi [\mnras] {10.1093/mnras/stx1400},
  \href {https://ui.adsabs.harvard.edu/abs/2017MNRAS.471.2224N} {471, 2224}

\bibitem[\protect\citeauthoryear{{Oliphant}}{{Oliphant}}{2007}]{oliphant2007python}
{Oliphant} T.~E.,  2007, \mn@doi [Computing in Science and Engineering]
  {10.1109/MCSE.2007.58}, \href
  {https://ui.adsabs.harvard.edu/abs/2007CSE.....9c..10O} {9, 10}

\bibitem[\protect\citeauthoryear{{Osato}, {Nishimichi}, {Oguri}, {Takada}  \&
  {Okumura}}{{Osato} et~al.}{2018}]{Osato2018}
{Osato} K.,  {Nishimichi} T.,  {Oguri} M.,  {Takada} M.,   {Okumura} T.,  2018,
  \mn@doi [\mnras] {10.1093/mnras/sty762}, \href
  {https://ui.adsabs.harvard.edu/abs/2018MNRAS.477.2141O} {477, 2141}

\bibitem[\protect\citeauthoryear{{Paz}, {Lambas}, {Padilla}  \&
  {Merch{\'a}n}}{{Paz} et~al.}{2006}]{paz2006shapes}
{Paz} D.~J.,  {Lambas} D.~G.,  {Padilla} N.,   {Merch{\'a}n} M.,  2006, \mn@doi
  [\mnras] {10.1111/j.1365-2966.2005.09934.x}, \href
  {https://ui.adsabs.harvard.edu/abs/2006MNRAS.366.1503P} {366, 1503}

\bibitem[\protect\citeauthoryear{{Pillepich} et~al.,}{{Pillepich}
  et~al.}{2018}]{pillepich2017first}
{Pillepich} A.,  et~al., 2018, \mn@doi [\mnras] {10.1093/mnras/stx3112}, \href
  {https://ui.adsabs.harvard.edu/abs/2018MNRAS.475..648P} {475, 648}

\bibitem[\protect\citeauthoryear{{Planck Collaboration} et~al.,}{{Planck
  Collaboration} et~al.}{2016}]{ade2016planck}
{Planck Collaboration} et~al., 2016, \mn@doi [\aap]
  {10.1051/0004-6361/201525830}, \href
  {https://ui.adsabs.harvard.edu/abs/2016A&A...594A..13P} {594, A13}

\bibitem[\protect\citeauthoryear{{Press} \& {Schechter}}{{Press} \&
  {Schechter}}{1974}]{press1974formation}
{Press} W.~H.,  {Schechter} P.,  1974, \mn@doi [\apj] {10.1086/152650}, \href
  {https://ui.adsabs.harvard.edu/abs/1974ApJ...187..425P} {187, 425}

\bibitem[\protect\citeauthoryear{{Richings}, {Frenk}, {Jenkins}, {Robertson}
  \& {Schaller}}{{Richings} et~al.}{2021}]{richings2021lens}
{Richings} J.,  {Frenk} C.,  {Jenkins} A.,  {Robertson} A.,   {Schaller} M.,
  2021, \mn@doi [\mnras] {10.1093/mnras/staa4013}, \href
  {https://ui.adsabs.harvard.edu/abs/2021MNRAS.501.4657R} {501, 4657}

\bibitem[\protect\citeauthoryear{{S{\'a}nchez}, {Baugh}, {Percival}, {Peacock},
  {Padilla}, {Cole}, {Frenk}  \& {Norberg}}{{S{\'a}nchez}
  et~al.}{2006}]{sanchez2006final}
{S{\'a}nchez} A.~G.,  {Baugh} C.~M.,  {Percival} W.~J.,  {Peacock} J.~A.,
  {Padilla} N.~D.,  {Cole} S.,  {Frenk} C.~S.,   {Norberg} P.,  2006, \mn@doi
  [\mnras] {10.1111/j.1365-2966.2005.09833.x}, \href
  {https://ui.adsabs.harvard.edu/abs/2006MNRAS.366..189S} {366, 189}

\bibitem[\protect\citeauthoryear{{Schaye} et~al.,}{{Schaye}
  et~al.}{2015}]{schaye2015eagle}
{Schaye} J.,  et~al., 2015, \mn@doi [\mnras] {10.1093/mnras/stu2058}, \href
  {https://ui.adsabs.harvard.edu/abs/2015MNRAS.446..521S} {446, 521}

\bibitem[\protect\citeauthoryear{{Schneider}, {Smith}  \& {Reed}}{{Schneider}
  et~al.}{2013}]{Schneider2013}
{Schneider} A.,  {Smith} R.~E.,   {Reed} D.,  2013, \mn@doi [\mnras]
  {10.1093/mnras/stt829}, \href
  {https://ui.adsabs.harvard.edu/abs/2013MNRAS.433.1573S} {433, 1573}

\bibitem[\protect\citeauthoryear{{Seljak}}{{Seljak}}{2000}]{seljak2000analytical}
{Seljak} U.,  2000, \mn@doi [\mnras] {10.1046/j.1365-8711.2000.03715.x}, \href
  {https://ui.adsabs.harvard.edu/abs/2000MNRAS.318..203S} {318, 203}

\bibitem[\protect\citeauthoryear{{Sheth} \& {Tormen}}{{Sheth} \&
  {Tormen}}{1999}]{sheth1999bias}
{Sheth} R.~K.,  {Tormen} G.,  1999, \mn@doi [\mnras]
  {10.1046/j.1365-8711.1999.02692.x}, \href
  {https://ui.adsabs.harvard.edu/abs/1999MNRAS.308..119S} {308, 119}

\bibitem[\protect\citeauthoryear{{Sheth} \& {Tormen}}{{Sheth} \&
  {Tormen}}{2002}]{sheth2002excursion}
{Sheth} R.~K.,  {Tormen} G.,  2002, \mn@doi [\mnras]
  {10.1046/j.1365-8711.2002.04950.x}, \href
  {https://ui.adsabs.harvard.edu/abs/2002MNRAS.329...61S} {329, 61}

\bibitem[\protect\citeauthoryear{{Smith} et~al.,}{{Smith}
  et~al.}{2003}]{smith2003clustering}
{Smith} R.~E.,  et~al., 2003, \mn@doi [\mnras]
  {10.1046/j.1365-8711.2003.06503.x}, \href
  {https://ui.adsabs.harvard.edu/abs/2003MNRAS.341.1311S} {341, 1311}

\bibitem[\protect\citeauthoryear{{Springel}}{{Springel}}{2010}]{springel2010arepo}
{Springel} V.,  2010, \mn@doi [\mnras] {10.1111/j.1365-2966.2009.15715.x},
  \href {https://ui.adsabs.harvard.edu/abs/2010MNRAS.401..791S} {401, 791}

\bibitem[\protect\citeauthoryear{{Springel}, {White}, {Tormen}  \&
  {Kauffmann}}{{Springel} et~al.}{2001}]{springel2001populating}
{Springel} V.,  {White} S. D.~M.,  {Tormen} G.,   {Kauffmann} G.,  2001,
  \mn@doi [\mnras] {10.1046/j.1365-8711.2001.04912.x}, \href
  {https://ui.adsabs.harvard.edu/abs/2001MNRAS.328..726S} {328, 726}

\bibitem[\protect\citeauthoryear{{Springel} et~al.,}{{Springel}
  et~al.}{2018}]{springel2018first}
{Springel} V.,  et~al., 2018, \mn@doi [\mnras] {10.1093/mnras/stx3304}, \href
  {https://ui.adsabs.harvard.edu/abs/2018MNRAS.475..676S} {475, 676}

\bibitem[\protect\citeauthoryear{{Tegmark} et~al.,}{{Tegmark}
  et~al.}{2004}]{tegmark2004sdss}
{Tegmark} M.,  et~al., 2004, \mn@doi [\apj] {10.1086/382125}, \href
  {https://ui.adsabs.harvard.edu/abs/2004ApJ...606..702T} {606, 702}

\bibitem[\protect\citeauthoryear{{Vega-Ferrero}, {Yepes}  \&
  {Gottl{\"o}ber}}{{Vega-Ferrero} et~al.}{2017}]{vega2017shape}
{Vega-Ferrero} J.,  {Yepes} G.,   {Gottl{\"o}ber} S.,  2017, \mn@doi [\mnras]
  {10.1093/mnras/stx282}, \href
  {https://ui.adsabs.harvard.edu/abs/2017MNRAS.467.3226V} {467, 3226}

\bibitem[\protect\citeauthoryear{{Vegetti}, {Koopmans}, {Bolton}, {Treu}  \&
  {Gavazzi}}{{Vegetti} et~al.}{2010}]{vegetii2010detection}
{Vegetti} S.,  {Koopmans} L.~V.~E.,  {Bolton} A.,  {Treu} T.,   {Gavazzi} R.,
  2010, \mn@doi [\mnras] {10.1111/j.1365-2966.2010.16865.x}, \href
  {https://ui.adsabs.harvard.edu/abs/2010MNRAS.408.1969V} {408, 1969}

\bibitem[\protect\citeauthoryear{{Vegetti}, {Koopmans}, {Auger}, {Treu}  \&
  {Bolton}}{{Vegetti} et~al.}{2014}]{vegetti2014inference}
{Vegetti} S.,  {Koopmans} L.~V.~E.,  {Auger} M.~W.,  {Treu} T.,   {Bolton}
  A.~S.,  2014, \mn@doi [\mnras] {10.1093/mnras/stu943}, \href
  {https://ui.adsabs.harvard.edu/abs/2014MNRAS.442.2017V} {442, 2017}

\bibitem[\protect\citeauthoryear{{Vogelsberger} et~al.,}{{Vogelsberger}
  et~al.}{2014}]{vogelsberger2014illustris}
{Vogelsberger} M.,  et~al., 2014, \mn@doi [\mnras] {10.1093/mnras/stu1536},
  \href {https://ui.adsabs.harvard.edu/abs/2014MNRAS.444.1518V} {444, 1518}

\bibitem[\protect\citeauthoryear{{Vogelsberger}, {Marinacci}, {Torrey}  \&
  {Puchwein}}{{Vogelsberger} et~al.}{2020}]{vogelsberger2020sims}
{Vogelsberger} M.,  {Marinacci} F.,  {Torrey} P.,   {Puchwein} E.,  2020,
  \mn@doi [Nature Reviews Physics] {10.1038/s42254-019-0127-2}, \href
  {https://ui.adsabs.harvard.edu/abs/2020NatRP...2...42V} {2, 42}

\bibitem[\protect\citeauthoryear{{Warren}, {Quinn}, {Salmon}  \&
  {Zurek}}{{Warren} et~al.}{1992}]{warren1992shapes}
{Warren} M.~S.,  {Quinn} P.~J.,  {Salmon} J.~K.,   {Zurek} W.~H.,  1992,
  \mn@doi [\apj] {10.1086/171937}, \href
  {https://ui.adsabs.harvard.edu/abs/1992ApJ...399..405W} {399, 405}

\bibitem[\protect\citeauthoryear{{Weinberger} et~al.,}{{Weinberger}
  et~al.}{2017}]{weinberger2017tng}
{Weinberger} R.,  et~al., 2017, \mn@doi [\mnras] {10.1093/mnras/stw2944}, \href
  {https://ui.adsabs.harvard.edu/abs/2017MNRAS.465.3291W} {465, 3291}

\bibitem[\protect\citeauthoryear{{Weiner}, {Serjeant}  \& {Sedgwick}}{{Weiner}
  et~al.}{2020}]{weiner2020RST}
{Weiner} C.,  {Serjeant} S.,   {Sedgwick} C.,  2020, \mn@doi [Research Notes of
  the American Astronomical Society] {10.3847/2515-5172/abc4ea}, \href
  {https://ui.adsabs.harvard.edu/abs/2020RNAAS...4..190W} {4, 190}

\bibitem[\protect\citeauthoryear{{Wetzel}, {Hopkins}, {Kim},
  {Faucher-Gigu{\`e}re}, {Kere{\v{s}}}  \& {Quataert}}{{Wetzel}
  et~al.}{2016}]{wetzel2016reconciling}
{Wetzel} A.~R.,  {Hopkins} P.~F.,  {Kim} J.-h.,  {Faucher-Gigu{\`e}re} C.-A.,
  {Kere{\v{s}}} D.,   {Quataert} E.,  2016, \mn@doi [\apjl]
  {10.3847/2041-8205/827/2/L23}, \href
  {https://ui.adsabs.harvard.edu/abs/2016ApJ...827L..23W} {827, L23}

\bibitem[\protect\citeauthoryear{{Xu} et~al.,}{{Xu}
  et~al.}{2009}]{xu2009aquarius}
{Xu} D.~D.,  et~al., 2009, \mn@doi [\mnras] {10.1111/j.1365-2966.2009.15230.x},
  \href {https://ui.adsabs.harvard.edu/abs/2009MNRAS.398.1235X} {398, 1235}

\bibitem[\protect\citeauthoryear{{Xu}, {Mao}, {Cooper}, {Gao}, {Frenk},
  {Angulo}  \& {Helly}}{{Xu} et~al.}{2012}]{xu2012effects}
{Xu} D.~D.,  {Mao} S.,  {Cooper} A.~P.,  {Gao} L.,  {Frenk} C.~S.,  {Angulo}
  R.~E.,   {Helly} J.,  2012, \mn@doi [\mnras]
  {10.1111/j.1365-2966.2012.20484.x}, \href
  {https://ui.adsabs.harvard.edu/abs/2012MNRAS.421.2553X} {421, 2553}

\bibitem[\protect\citeauthoryear{{Xu}, {Sluse}, {Gao}, {Wang}, {Frenk}, {Mao},
  {Schneider}  \& {Springel}}{{Xu} et~al.}{2015}]{xu2015flux}
{Xu} D.,  {Sluse} D.,  {Gao} L.,  {Wang} J.,  {Frenk} C.,  {Mao} S.,
  {Schneider} P.,   {Springel} V.,  2015, \mn@doi [\mnras]
  {10.1093/mnras/stu2673}, \href
  {https://ui.adsabs.harvard.edu/abs/2015MNRAS.447.3189X} {447, 3189}

\bibitem[\protect\citeauthoryear{{Yoon}, {Johnston}  \& {Hogg}}{{Yoon}
  et~al.}{2011}]{yoon2011streams}
{Yoon} J.~H.,  {Johnston} K.~V.,   {Hogg} D.~W.,  2011, \mn@doi [\apj]
  {10.1088/0004-637X/731/1/58}, \href
  {https://ui.adsabs.harvard.edu/abs/2011ApJ...731...58Y} {731, 58}

\bibitem[\protect\citeauthoryear{{Zentner}, {Kravtsov}, {Gnedin}  \&
  {Klypin}}{{Zentner} et~al.}{2005}]{zentner2005anis}
{Zentner} A.~R.,  {Kravtsov} A.~V.,  {Gnedin} O.~Y.,   {Klypin} A.~A.,  2005,
  \mn@doi [\apj] {10.1086/431355}, \href
  {https://ui.adsabs.harvard.edu/abs/2005ApJ...629..219Z} {629, 219}

\bibitem[\protect\citeauthoryear{{Zhu}, {Marinacci}, {Maji}, {Li}, {Springel}
  \& {Hernquist}}{{Zhu} et~al.}{2016}]{zhu2016baryonic}
{Zhu} Q.,  {Marinacci} F.,  {Maji} M.,  {Li} Y.,  {Springel} V.,   {Hernquist}
  L.,  2016, \mn@doi [\mnras] {10.1093/mnras/stw374}, \href
  {https://ui.adsabs.harvard.edu/abs/2016MNRAS.458.1559Z} {458, 1559}

\bibitem[\protect\citeauthoryear{{van der Walt}, {Colbert}  \&
  {Varoquaux}}{{van der Walt} et~al.}{2011}]{van2011numpy}
{van der Walt} S.,  {Colbert} S.~C.,   {Varoquaux} G.,  2011, \mn@doi
  [Computing in Science and Engineering] {10.1109/MCSE.2011.37}, \href
  {https://ui.adsabs.harvard.edu/abs/2011CSE....13b..22V} {13, 22}

\makeatother
\end{thebibliography}

\appendix

\section{Counting statistics}

\subsection{Sampling variance}
\label{sec:dmo.sampling}
\begin{figure*}
    \centering
    \includegraphics[width=0.925\textwidth]{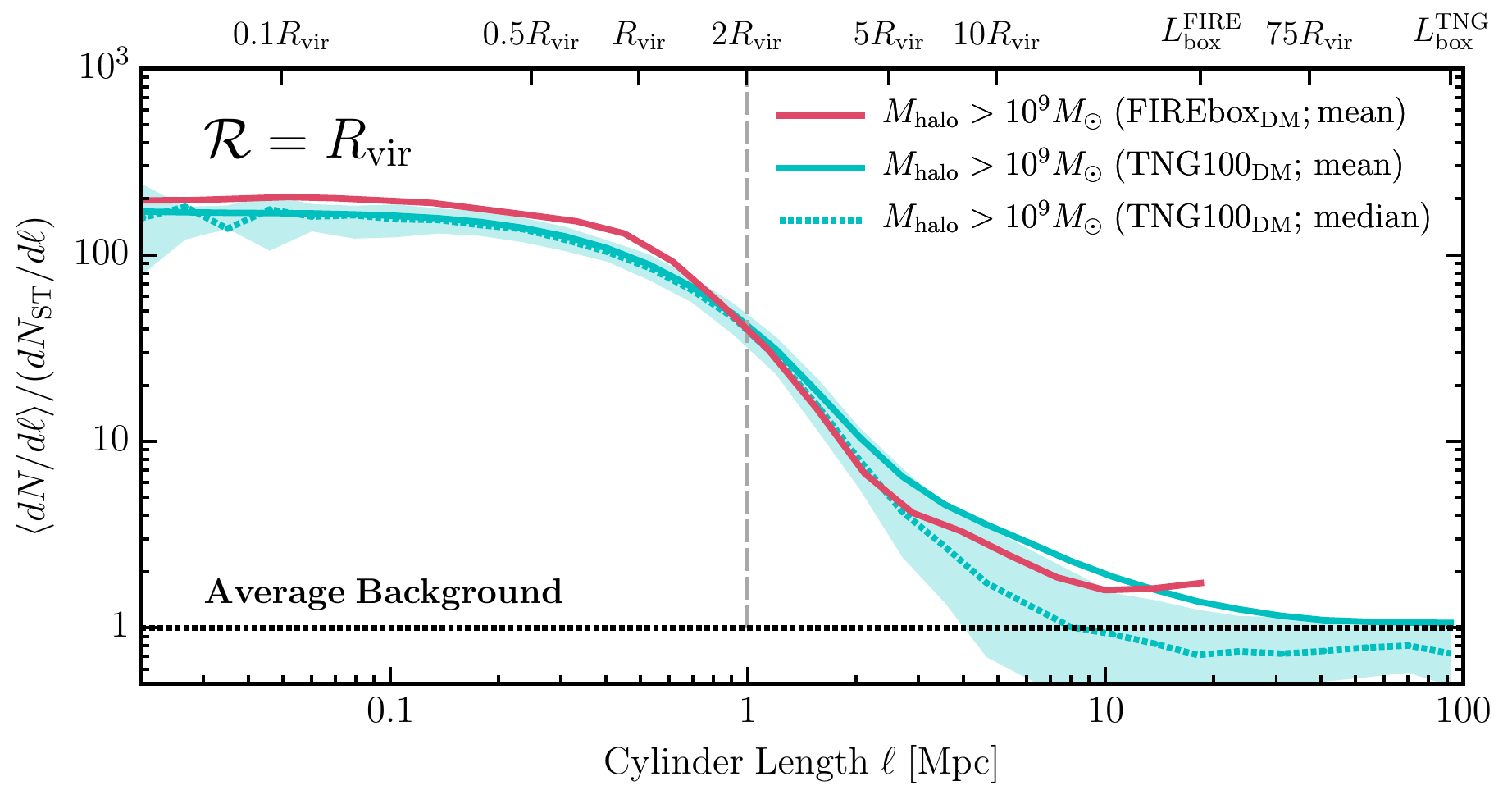}
    \caption{
    Analogous to Fig.~\ref{fig:4}, now with results of \tngdm{} (cyan) and \fbxdm{} (red) for $M_{\rm halo} >$ \mNine{} for cylinders with a radius size comparable to the lens mass halo ($\mathcal{R} = R_{\rm vir}$) as opposed to the lens halo's Einstein radius ($\mathcal{R} = 10\,{\rm kpc}$). The solid curves depict the \textit{mean} differential count while the dashed cyan is the {\emph median} differential count for \tngdm{}. The cyan band encloses the $\pm 1\sigma$ region of the curves. The median counts fall below the mean counts, indicating that the mean counts are strongly affected by rare, high-count orientations.
    }
    \label{fig:A1}
\end{figure*}

In Section~\ref{sec:rate.of.counts}, we discussed the possibility that the differences in local clustering signal between \tngdm{} and \fbxdm{} could arise from sampling variance -- there are only three lens-mass hosts in \fbxdm{} -- rather than differences in large-scale structure (compare the red and cyan curve in Fig.~\ref{fig:4}). 
In order to explore this more fully, Fig.~\ref{fig:A1} shows $\langle dN/d\ell \rangle$ for \tngdm{} (cyan) and \fbxdm{} (red). This figure mirrors Fig. \ref{fig:4} except now we are using a cylinder radius with the size of the lens-mass virial radius ($\mathcal{R} = R_{\rm vir}\simeq 500$ kpc) as opposed to the typical Einstein radius ($\mathcal{R} = 10$ kpc) in order to improve counting statistics. The solid lines depict the mean of all projections while the dashed (shown only for \tngdm) is the {\em median} of all projections. The cyan band encloses the $\pm 1\sigma$ of all of the individual curves around median. 

For $\ell < 1$ Mpc, the median and mean curve for \tngdm{} are consistent with one another, but for $\ell > 1$ Mpc we see that the average count (which is what we plot in  Fig.~\ref{fig:4}) sits well above the median. This is indicative of a highly non-Gaussian distribution, with a tail skewed towards large fluctuations, as expected for non-linear dark matter structure. Specifically, rare (high-count) events drive the average higher than the median. If this is the case, we require a large number of realizations in order to sample enough of the distribution to capture the true average. As further evidence that sample variance is the cause of the differences in local clustering signal between \tngdm{} and \fbxdm, the median line in Fig.~\ref{fig:A1} falls {\em below} the average background (black dashed) at $\ell \approx 10$ Mpc. This is due to a large number of null counts.

Though the {\em average} count from \fbxdm{} falls below the average line from \tngdm{}, we see that it falls within the $1\sigma$ band about the {\em median}. Given that we only have three host halos in our sample, this would not be unexpected even if the halos were sampling the same large-scale structure field. Naively, there is a $\sim 30 \%$ chance that three randomly drawn distributions will lie within $1\sigma$ of the median. Given that we only have three host halos, we conclude that the observed result is consistent with small-number statistics. 

\begin{figure*}
    \centering
    \includegraphics[width=0.925\textwidth]{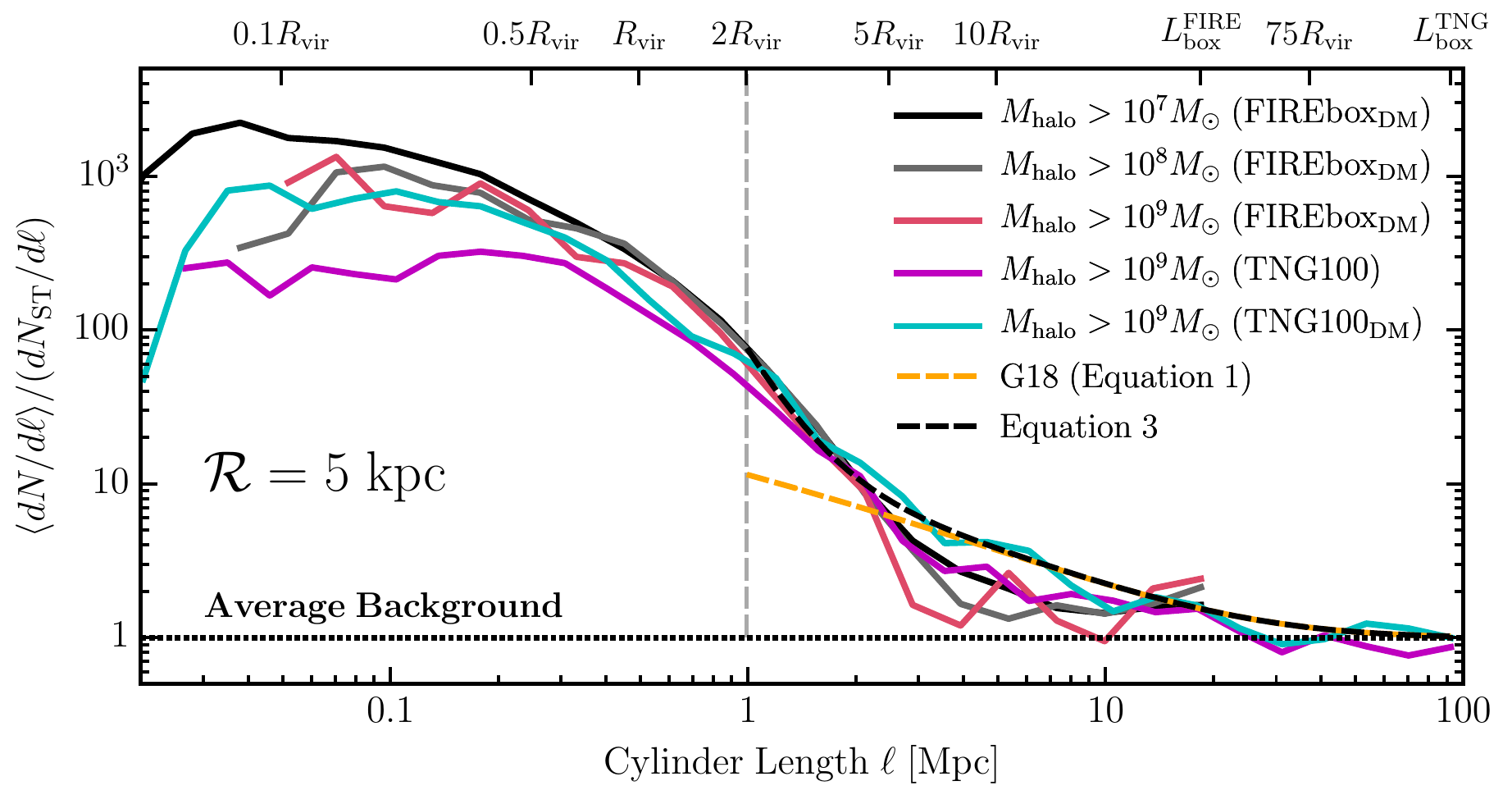}
    \includegraphics[width=0.925\textwidth]{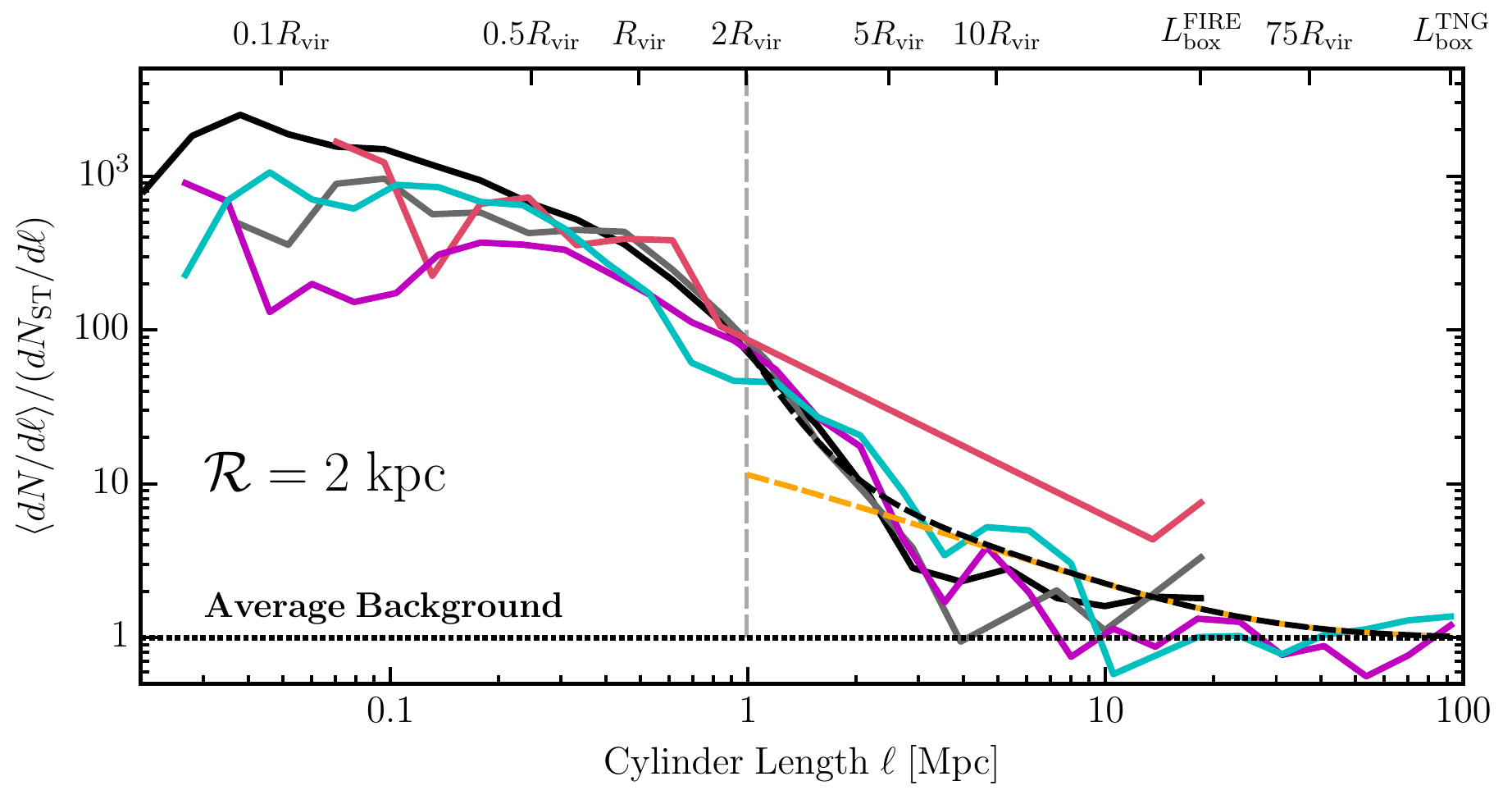}
    \caption{
    Analogous to Fig.~\ref{fig:4}, now for cylinders with a radius size $\mathcal{R} = 5$ kpc and a much smaller annulus $\mathcal{R} = 2$ kpc. While the curves are much noisier as the radius size becomes smaller (owing to the increased fraction of null sight lines), the we are able to recover the same trend for the differential counts seen for a projected radius of $\mathcal{R} = 10$ kpc.
    }
    \label{fig:A2}
\end{figure*}

\subsection{Choice of cylinder radius}
\label{sec:cylinder.choice}
The conclusions made in the main text are based on the average number density of subhalos within a project cylinder radius of 10 kpc, which is comparable to the typical Einstein ring of our $10^{13}\ M_{\odot}$ lens-mass halos. 
The choice of $\mathcal{R}=10\rm\, kpc$ could bias our results towards higher substructure, and potentially, LOS halo counts, as massive galaxy lenses typically have Einstein ring radii of $\sim 5$ kpc \citep{bolton2008sloan}.

In Fig.~\ref{fig:A2}, we demonstrate how the selection of smaller lens radii does not impact the trends discussed in our main analysis. Both the top and bottom plot mirrors exactly Fig.~\ref{fig:4}, now with the top and bottom plot showing results for a projected radius of 5 and 2 kpc, respectively. We find that we are able to mostly recover the same trend seen for the 10 kpc projected radius from Fig.~\ref{fig:4} for both $\mathcal{R}=$ 5 and 2 kpc.
In the 5 kpc radius projection, the curves are more noisier than the 10 kpc owing to the increasing fraction of projections with zero halos. We would argue that the $\mathcal{R}=5$ kpc size projections have comparable trends to the $\mathcal{R}=10$ kpc case. Out to $\ell=5 R_{\rm vir}$, most of the mean differentials drives faster down to the average compared to the 10 kpc case, but we suspect this is owed to low number statistics along with the sampling variance for \tngdm{} (see discussion in Appendix~\ref{sec:dmo.sampling}). This becomes more apparent for the smaller projected radius of 2 kpc, as the fraction of projections of zero halos increases appreciably. Note that the $M_{\rm halo}>$\mNine{} result is impacted greatly by sampling variance the low number of halos along the LOS.

\section{Supplementary Discussion with Dark Matter Only Physics}
\label{sec:dmo.discussion}

This appendix presents the DMO results from \tngdm{} and for \fbxdm and provides additional elaboration of the results in the main text. 
Section~\ref{sec:dmo.excess} picks up from later discussion of Section~\ref{sec:counts} and  Section~\ref{sec:dmo.principal} picks up from Section~\ref{sec:principal}.

\begin{figure}
    \centering
    \includegraphics[width=\columnwidth]{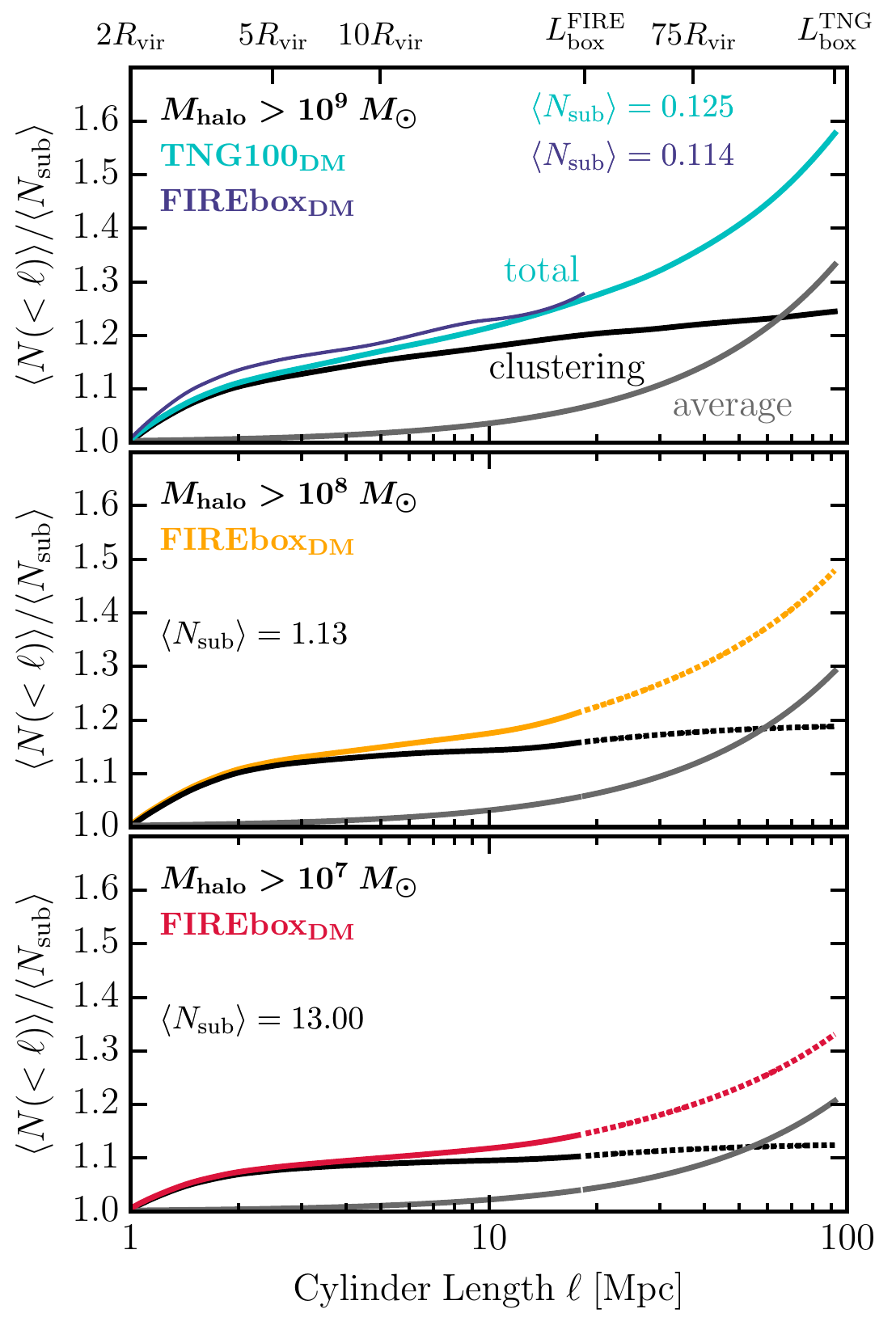}
    \caption{
    Like Fig.~\ref{fig:7}, now showing only the results for \tngdm{} and \fbxdm. In the top panel, the black and gray curve are computed based off of the total counts from \tngdm{} (cyan curve). As a comparison, the blue curve shows the total counts from \fbxdm{} out to $L^{\rm FIRE}_{\rm box}$. The middle and bottom panel are presented similarly with their curves extrapolated out to the $L^{\rm TNG}_{\rm box}$ using Eq.~\eqref{eq:analytical.los.corrected}.
    }
    \label{fig:B1}
\end{figure}

\subsection{Clustering contribution to the line-of-sight population}
\label{sec:dmo.excess}
Fig.~\ref{fig:B1} presents the contribution of the simulated clustering to the LOS halo population in \tngdm{} and \fbxdm. Namely, the top, middle, and bottom panels plots the subhalo populations based on the lower-mass cuts of $10^{9}\ M_{\odot}$, $10^{8}\ M_{\odot}$, and $10^{7}\ M_{\odot}$, respectively. The colored curves, labeled ``total'', are the actual results of the simulation out to the box using our method discussed in the main text. The average background expectation for halos above the low-mass cuts in each panel is depicted by the grey curves. The clustering contribution from the simulations, labeled ``clustering'', are plotted as the black curves, which is the difference between the ``total'' curve and ``average'' curves. Like before, all results are normalized by the $\langle N_{\rm sub }\rangle$ quantified from out lens-mass sample. Since the actual \fbxdm{} results extend out to \lfbx, we extrapolate the curves out to \ltng using Eq.~\eqref{eq:analytical.los} (using Eq.~\ref{eq:analytical.los.corrected} is also equally adequate).

\begin{figure}
    \centering
    \includegraphics[width=\columnwidth]{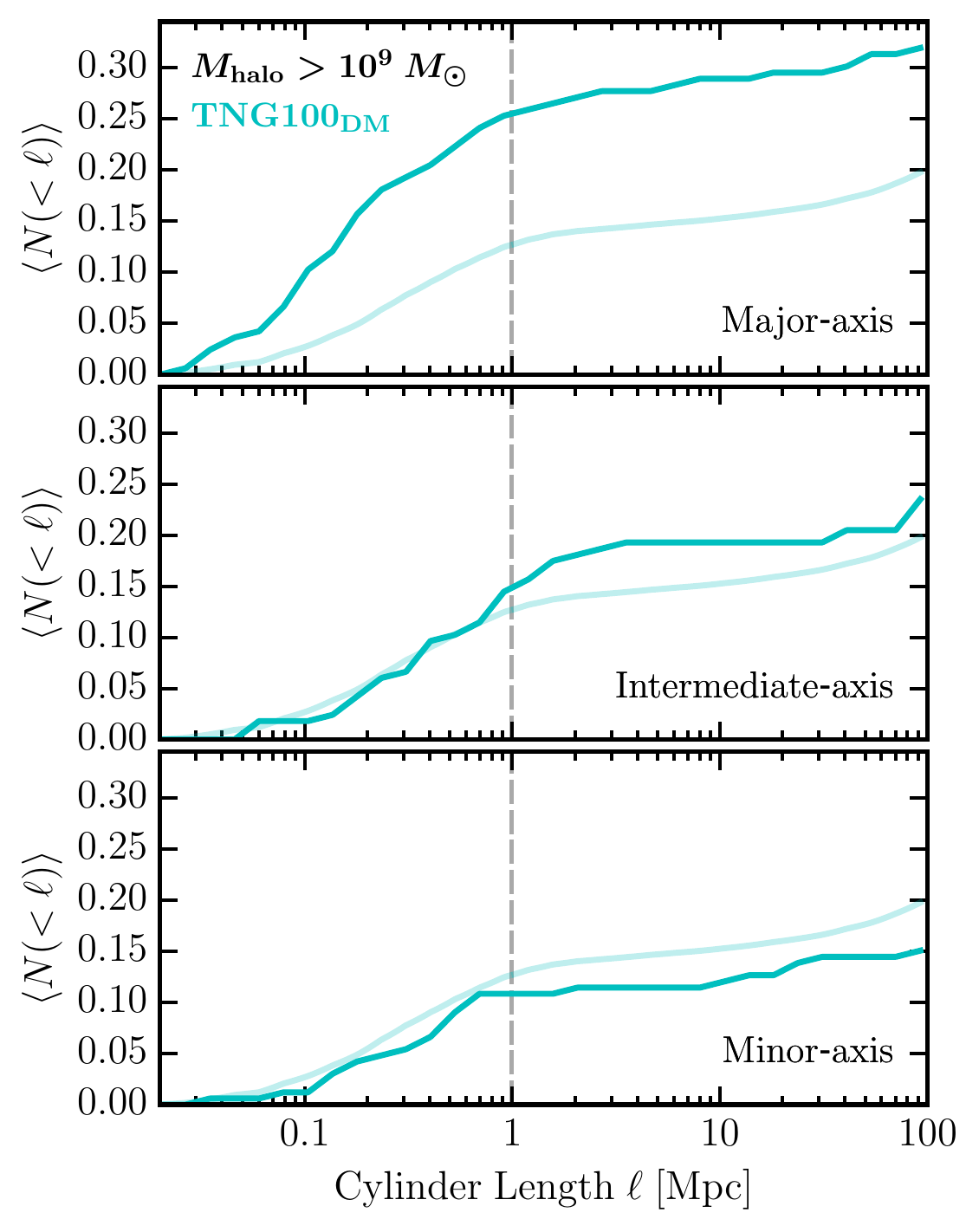}
    \caption{
        Like Fig.~\ref{fig:8}, but now showing the \tngdm{} counts for halos $M_{\rm halo} > 10^{9}\ M_{\odot}$ along the major, intermediate, and minor axis in the top, center, and bottom panels, respectively. Plotted for comparison, the transparent lines are the mean counts of the substructure with local clustering counts for \tngdm{} presented previously in Fig.~\ref{fig:5}.
    }
    \label{fig:B2}
\end{figure}

Starting with the counts $10^{9}\ M_{\odot}$ (top panel), clustering acts as the most contributing component to the LOS halo population until the average background takes over at $\ell \approx 75R_{\rm vir}$ (or $r\approx 37.5\ \rm Mpc$) for \tngdm{} (cyan curve). As a comparison check, we plotted the \fbxdm{} results of $10^{9}\ M_{\odot}$ (thin blue curve) and find minimal difference in results. Note the clustering component is determined from \tngdm{} and not \fbxdm. The clustering component in \tngdm{} boosts about $1.5\langle N_{\rm sub} \rangle$ out until the average background takes over at $\ell\approx 70\ \rm Mpc$ (or $r\approx35$ Mpc). This is less significant than the \tng{} in the main text, the clustering component strongly boosts the number of LOS halos to about $1.7\langle N_{\rm sub} \rangle$ until the average background takes over.

As we go further down to the low-mass cuts of the subhalo populations, the contribution from correlated clustering becomes much weaker for decrease mass. For $M_{\rm halo} > 10^{8}\ \rm M_{\odot}$ (middle panel), the clustering component boosts the number of LOS halos to only about $\sim 1.2 \langle N_{\rm sub}\rangle$ until the average background takes over at about $\ell \approx 60\ \rm Mpc$ (or $r\approx30$ Mpc). The clustering component for $M_{\rm halo} > 10^{7}\ \rm M_{\odot}$ (bottom panel) becomes weaker by only boosting the number of LOS halos to only about $\sim 1.1 \langle N_{\rm sub}\rangle$ until the average background takes over at $r \approx 30$ Mpc). In order to conclude with more robust predictions, a larger sample lens-mass halos in comparable cosmological environments is needed to reduced the uncertainty possibly found in $\langle N_{\rm sub}\rangle$ for these low-mass halos.

\subsection{Structure along principal axes}
\label{sec:dmo.principal}
Fig.~\ref{fig:B2} plots the mean subhalo counts as seen from lens-centered projections along the principal axis of the \tngdm{} lens-mass halos. The axes were computed using the method detailed in Section~\ref{sec:principal}. Shown are only the results from \tngdm{} since these cosmological volumes have enough lens-centered mass halos to provide enough statistics since each halo has only three principal axis to orientate on. Moreover, \fbxdm, while useful for simulating $10^{7-8}\ M_{\odot}$ halos in cosmological environment, has only three lens-centered hosts, which will not provide adequate statistics to present. Though, we find similar trends. The top, middle, and bottom panels depict the average counts along the major-, intermediate-, and minor-axis, respectively. For comparison, the faded solid line in each panel plots the mean counts presented previously in Fig.~\ref{fig:5}. We again see that projections along the major-axis of DMO lens-mass halos, subhalos tend to populate on the densest principal axis. This effect is more dramatic for the substructure compared to the \tng{} shown previously, owing to the lack of a central galaxy.

Projections the size of the Einstein radius along the major-axis find around $75\%$ of the projections to contain no substructure out to the size of the halo. Along the minor-axis, this is around $\sim 90\%$. Additionally, the substructure counts appear to be almost comparable to the mean from Fig.~\ref{fig:4}, though slightly less. Done on the intermediate-axis, this results in counts somewhat comparable to the mean counts, though we see ${\sim}5\%$ boost in the LOS component.

\section{Comparison between mass functions of lens-mass sample}
\label{sec:compare}
As mentioned in Section~\ref{sec:numerical.methods}, the \subfind{} halo finder was ran for the TNG suites while \rockstar{} was applied to \fbxdm. Fig.~\ref{fig:1} provides confidence in the resulting halo catalogs used in the main text, as both catalogs show robust agreement with the analytical prediction. However, halo finders vary in routines for quantifying the masses for subhalos and can potentially produce different mass functions when ran to the same simulations. It would be worth comparing our resulting halo finders with one another based on the resolved substructure population found for our lens-target halos at $z=0.2$.

\begin{figure*}
    \centering
    \includegraphics[width=0.95\textwidth]{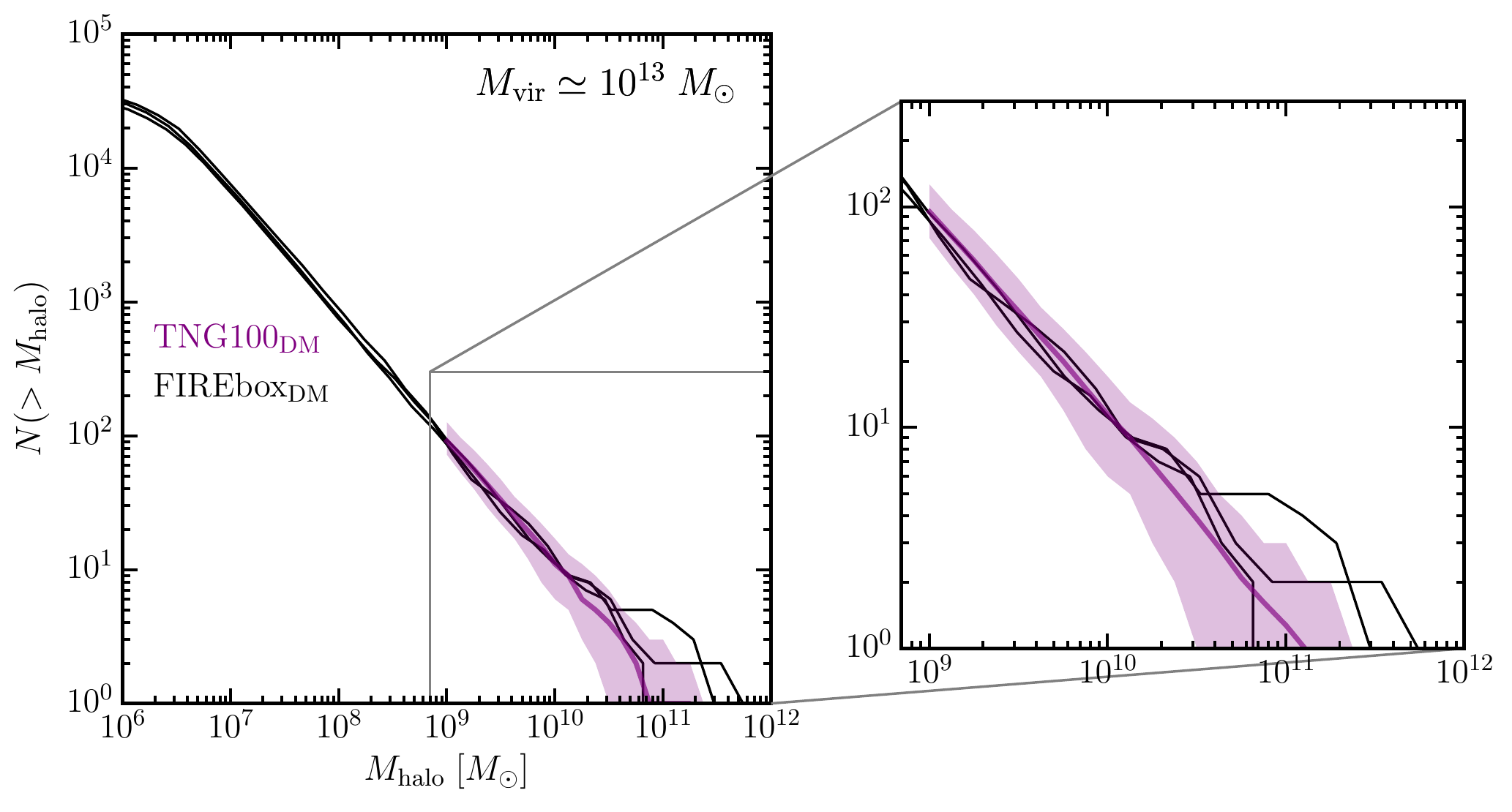}
    \caption{
    The subhalo mass functions for the lens-target halos in our main sample compared between the \tngdm{} (purple curve) and \fbxdm{} (black curves). \fbxdm{} follows the median \tngdm{} curves sufficiently for the subhalo masses ranging from $\sim 10^{9-10}\ M_{\odot}$ while the three lens-target in \fbxdm{} hosts several more massive $10^{11}\ M_{\odot}$. Though, the curves are still mostly found enclosed in the $90\%$ dispersion. The agreement depicted from the subhalo mass functions provides further justification for the choice of our {\small ROCKSTAR} parameters detailed in Section~\ref{sec:numerical.methods}.
    }
    \label{fig:C1}
\end{figure*}

\begin{figure*}
    \centering
    \includegraphics[width=0.95\textwidth]{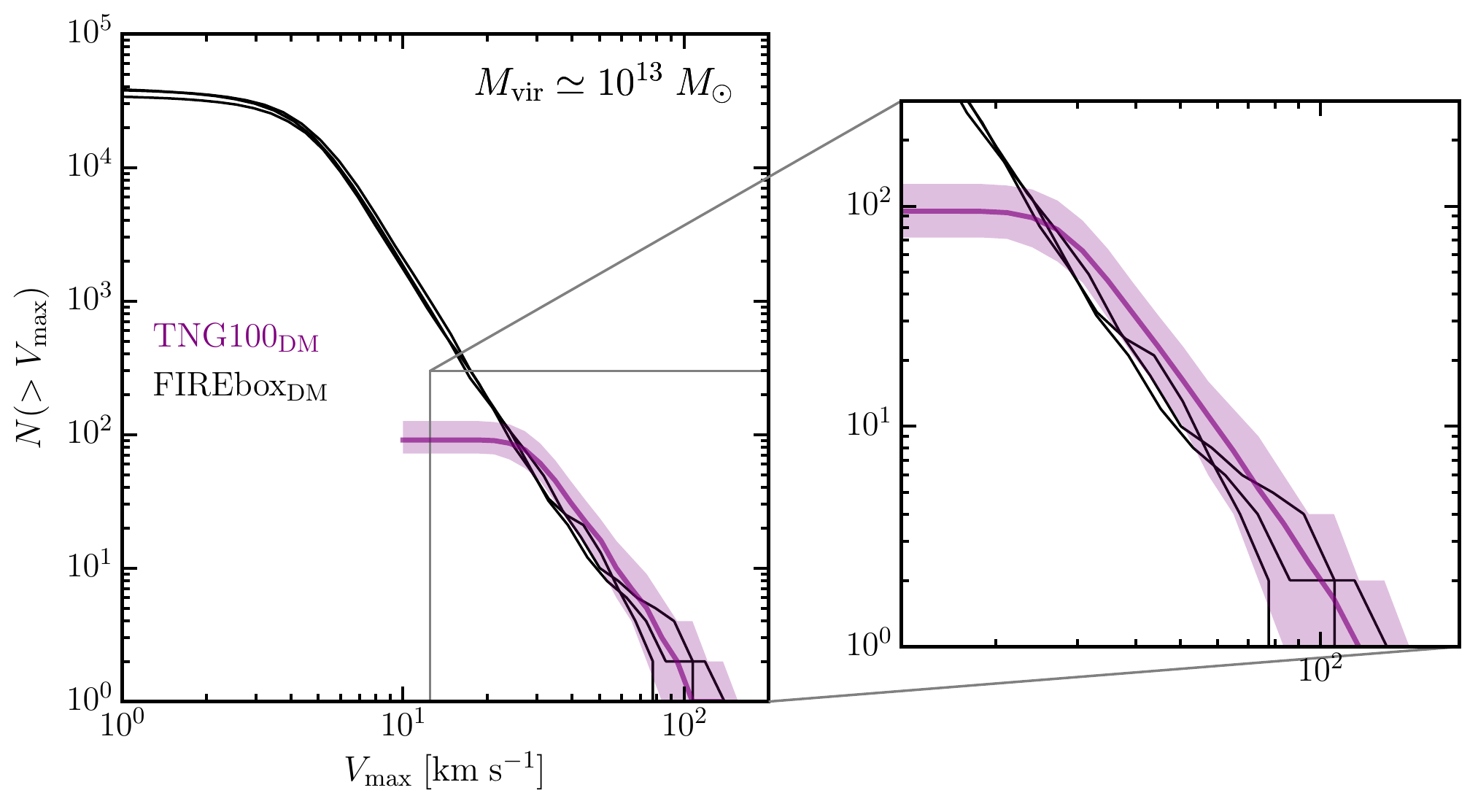}
    \caption{
    The subhalo $V_{\rm max}$ functions for the lens-target halos in our main sample compared between the \tngdm{} (purple curve) and \fbxdm{} (black curves).
    }
    \label{fig:C2}
\end{figure*}

Fig.~\ref{fig:C1} shows the subhalo mass functions for the $10^{13}\ M_{\odot}$ target-lens systems in \tngdm{} (purple) while the three from \fbxdm{} (black curves). The purple band for \tngdm{} encloses the $90\%$ dispersion of all of the target-lens subhalo mass functions.
Significant disagreement could arise based on the assignment of particles to subhalos from halo finders, \citep{graus2018smooth}. To clarify this point, Fig.~\ref{fig:C2} provides a different view of subhalo sample, where the subhalo $V_{\rm max}$ function for the same lens-target systems.

\bsp	
\label{lastpage}
\end{document}